\documentclass[12pt,preprint]{aastex}
\slugcomment{Accepted for publication in The Astrophysical Journal}
\shorttitle{Chromospheric Activity in M67}
\shortauthors{Giampapa et al.}
\begin{document}
\title{A Survey of Chromospheric Activity in the Solar-Type Stars in 
the Open Cluster M67\altaffilmark{1}}
\altaffiltext{1}{The results presented herein are based on data obtained at 
the WIYN telescope and at the McMath-Pierce Solar Telescope, 
respectively.  The WIYN Observatory is a joint facility of the University of 
Wisconsin-Madison, Indiana University, Yale University, and the
National Optical Astronomy Observatory.  The McMath-Pierce Solar Telescope
Facility is operated by the National Solar Observatory for the National 
Science Foundation}
\author{Mark S. Giampapa\altaffilmark{2}}
\altaffiltext{2}{National Solar Observatory/NOAO, 950 N. Cherry Ave., POB 26732, Tucson,
AZ 85726-6732. The NSO and the NOAO are each operated by the
Association of Universities for Research in Astronomy, Inc. (AURA) under
cooperative agreement with the National Science Foundation}
\email{giampapa@noao.edu}
\author{Jeffrey C. Hall\altaffilmark{3}}
\altaffiltext{3}{Lowell Observatory, 1400 W. Mars Hill Rd., Flagstaff, AZ
86001}
\email{jch@lowell.edu}
\author{Richard R. Radick\altaffilmark{4}}
\altaffiltext{4}{Air Force Research Laboratory, NSO/Sacramento Peak, Box
62, Sunspot, NM  88349}
\email{radick@nso.edu}
\and
\author{Sallie L. Baliunas\altaffilmark{5}} 
\altaffiltext{5}{Harvard-Smithsonian Center for Astrophysics, 60 Garden St., MS
15, Cambridge, MA  02138}
\email{baliunas@cfa.harvard.edu}

\begin{abstract}

We present the results of a spectroscopic survey of the Ca II H \& K core 
strengths
in a sample of 60 solar-type stars that are members of the solar-age and
solar-metallicity open cluster M67.  
We adopt the HK index, defined as the summed H+K core strengths 
in 1 {\AA} bandpasses centered on the H and K lines, respectively, 
as a measure of the chromospheric activity that is present.  
We compare the distribution of mean HK index values for the M67 
solar-type stars with the variation of this index as measured for the Sun 
during the contemporary solar cycle. 
We find that the stellar distribution in our HK
index is broader than that for the solar cycle.  
Approximately 17\% of the M67
sun-like stars exhibit average HK indices that are less than solar minimum.
About 7\%-12\% are
characterized by relatively high activity in excess of solar maximum
values while 72\%-80\% of the solar analogs exhibit Ca II H+K strengths
within the range of the modern solar cycle.  The ranges given reflect
uncertainties in the most representative value of the maximum
in the HK index to adopt for the solar cycle variations
observed during the period A.D. 1976---2004.
Thus, $\sim$ 20\%-30\% of our 
homogeneous sample of sun-like stars have mean 
chromospheric H+K strengths that are
outside the range of the contemporary solar cycle.  
Any cycle-like variability that is present in the M67 solar-type
stars appears to be characterized by periods greater than $\sim$ 6 years. 
Finally, we estimate a mean chromospheric age for M67 in the range of
3.8---4.3 Gyr.
\end{abstract}


\section{Introduction}

A key objective of investigations in the field of solar-stellar astrophysics
is to provide insight on the nature and possible range of solar magnetic
field-related atmospheric activity through observations of stellar 
analogs of the Sun.  In this way, we immediately obtain information on
the long-term variability of the Sun that otherwise would not be
possible (or practically feasible) with, for example, the modern 
solar data base of high-precision, chromospheric Ca II H and K observations 
of about three decades (e.g., White \& Livingston 1978; Livingston 1994).
This is especially important
given that the amplitude of solar and stellar variations in brightness
are usually highly correlated with cycle variations 
in chromospheric emission (Hudson 1988;Radick 1991; Fr\"{o}hlich 1994; de Toma
et al. 2004).  Thus, the range of cycle 
variability in a sample of solar-type stars can be indicative of the 
magnitude of total brightness changes (Zhang et al. 1994) as
well as the nature of variability in the spectral irradiance or the particle
fluence in the 
Sun.  These forms of solar variability, in turn, may affect global climate 
change (e.g., Soon et al. 1996; Soon et al. 2000a,b; Foukal 2003; Stott et al.
2003; Haigh 2003, 2005; 
Soon \& Yaskell 2004 and references therein;  Lean et al. 2005). 

In addition to the impact on our understanding of the role of the past 
and future variability of the Sun in global climate, the characteristics
of cycle-related variability are a manifestation of the nature of the
dynamo mechanism that is operative in the Sun and sun-like
stars (Saar \& Baliunas 1992; Saar 2002).  Furthermore, the calibration
of the empirical correlation between stellar age and chromospheric
activity (Walter \& Barry 1991; Soderblom et al. 1991; Simon 1992;
Donahue 1998) will be affected by the intrinsic dispersion
introduced by the cycle-related variation in chromospheric emission. 
As an illustration of the magnitude of this effect, if we combine the
25\% range in the Ca II index from minimum to maximum in the solar cycle,
as measured by Livingston \& Wallace (2003) from 1975 to 2003, with the
exponential decay law proposed by Walter \& Barry (1991) for Ca II
then the apparent uncertainty in the age of the Sun arising from its
cycle variability would be $\sim$ 21\%, or an age-spread of roughly 4 Gyr
-- 6 Gyr (see also Baliunas et al. 1998, their Fig. 4).

Recent studies of the comparative properties of the Sun and sun-like stars
have emphasized field stars in the solar neighborhood.
This approach yields samples that are accessible to detailed and
frequent observation with small-to-moderate aperture telescopes, and
includes the extensive archive of Ca II H and K observations from the
Mt. Wilson program (Wilson 1978; Vaughan et al. 1978; Baliunas et al.
1995).  Baliunas \& Jastrow (1990) utilized data from the Mt. Wilson
program to examine the nature and distribution of activity in 74 stars
that were selected to be solar-type according to a set of criteria
that included age, mean level of chromospheric activity, and mass as
inferred from photometric $B - V$ color.  They estimated 
the potential range of solar magnetic
activity over time scales of centuries, and the approximate magnitude of
the associated changes that would be implied in the solar irradiance, 
particularly in the regime of exceptionally low-activity that
is presumed to have occurred during the Maunder Minimum (White 1978). 
Among the distinctive features of the tentative results discussed by 
Baliunas \& Jastrow
(1990) were a significant width of the distribution of mean chromospheric 
activity measured at random phases of centuries-long variability and
a possible bimodal distribution of mean chromospheric activity.
They inferred from their stellar sample, using a particular method of sampling
the time series of observations, that
exceptionally quiescent levels of activity occurred at a frequency of about
one-third of the time.  This estimate, in turn, was primarily associated 
with four stars in their sample 
that exhibited little variability (i.e., instrumentally flat time series in 
the Mt. Wilson S-index of relative Ca II H and K strength) over periods of 
approximately two decades.
Baliunas \& Jastrow (1990) suggested that these objects were in the stellar
counterpart of the solar Maunder-minimum state.  In a subsequent study of 
a subsample of 10 solar-type stars, Zhang et al. (1994) cautioned that 
the small number of nearby solar-type stars, and the heterogeneity 
that is inherent in such field star samples, needed to be addressed.

Hall \& Lockwood (2004) reexamined the conclusions by Baliunas \&
Jastrow (1990) using a refined sample of solar counterparts in the field as
monitored since 1994 with their Solar-Stellar Spectrograph (SSS) at Lowell
Observatory in Flagstaff, Arizona (Hall \& Lockwood 1995). 
These investigators also find that
one-third of the solar counterparts in their sample do not exhibit
cyclic behavior, in agreement with the results of Baliunas \& Jastrow
(1990).  However, in contrast to the earlier study,  the 
distribution of H and K activity in the Hall \& Lockwood (2004) sample
is unimodal rather than bimodal.  Moreover, these investigators 
note that the level of
chromospheric H and K emission in the non-cyclic, flat stars spans a
range from levels below that of solar minimum to greater than that
seen at solar maximum.  Hence, as Hall \& Lockwood conclude,  non-cyclic 
behavior is not necessarily
associated with relatively reduced chromospheric activity.

These kinds of studies illustrate the crucial discrepancies in conclusions
that can arise from investigations of heterogeneous samples of field
stars.  For example, in the case of an ostensibly 
solar-type star that exhibits a 
low-level of activity, it is particularly 
critical to know its evolutionary status before concluding that    
it is a Maunder-minimum candidate.  In fact, Wright (2004) claims,
on the basis of an analysis of $Hipparcos$ data and a transformation of 
the Mt. Wilson S-index to a normalized chromospheric emission 
flux ($R^{\prime}_{HK}$), that nearly all the
stars thus far identified as Maunder minimum candidates are actually
evolved or subgiant stars with intrinsically low (and constant)
chromospheric activity rather than solar-type stars that are in a
temporary state of quiescence.  A further examination of this result based
on high resolution H \& K line spectra combined with an independent flux 
calibration technique (e.g., following the methods of Linsky et al. 1979) 
is merited. 

 In this investigation, we extend these pioneering studies to a sample
of solar-type cluster members that are homogeneous in age and chemical
composition.  The open cluster M67 is an especially appropriate
target of observation since it is approximately the same 
age (about 5 Gyr $\pm$ 1 Gyr; Demarque, Guenther \& Green 1992) and of the 
same metallicity as the Sun (Barry \& Cromwell 1974).  Thus, the
solar-type members represent analogs of the Sun at random
phases in their activity cycles.  Therefore, a survey of the ``Suns of
M67" should reveal the possible range of {\it solar} chromospheric
activity, thus immediately yielding information on the potential
long-term variability of the Sun.  We discuss in $\S2$ the observations and
reduction.  In $\S3$, we describe the methodology adopted for the 
calibration of the stellar spectra to relative intensity and discuss our
estimates of the random and systematic errors.  We
present the results in $\S4$ followed by discussion in $\S5$.  We
summarize our findings and indicate the future directions of this program
in $\S6$.

\section{Observations}

 We utilized the 3.5-m WIYN telescope in conjunction with the Hydra
multi-fiber positioner to perform multi-object spectroscopy over a
one-degree field. The Hydra instrument is described by Barden \& Armandroff
(1995).  
We used a bench-mounted spectrograph with the T2KC CCD,
Simmons camera, the blue fiber cables (3 arcsec diameter) and 
the 1200 $g$ mm$^{-1}$ grating in second order.
A CuSO$_{4}$ blocking filter was employed to suppress red leak.
This configuration yields a spectral resolution of 
0.80 {\AA} $\pm$ 0.03 {\AA}, as measured by the FWHMs of CuAr comparison 
lines extending from 3875 {\AA} to 3995 {\AA}.

The stars are selected from the Girard et al. (1989) proper motion
study combined with the CCD photometry given by Montgomery, Marschall \&
Janes (1993).  For those objects from the Girard et al. study
where photometry from
Montgomery et al. is not available, we adopt the photometry given by
Sanders (1989).   We note that Girard et 
al. determine two values of the membership probability:  One is  based 
solely on proper motion while the other value  
takes into account the position of a star
relative to the center of the cluster.  This latter membership probability
assumes a particular model for the spatial distribution of cluster stars.  
In general, we include stars with  membership probabilities 
that are $\gtrsim$ 90\% in one or the other of the 
probabilities given by Girard et al. (1989).  In a few cases, we 
include objects with membership probabilities that are slightly less than 
90\% in both estimates.  In these instances, the position of the object in 
the color-magnitude diagram provides further support
for the assumption that the star is a bona-fide cluster member. 

Our photometric criteria
include stars in the range of apparent brightness of approximately
14 $\leq~V~\lesssim$ 15
and intrinsic color of +0.58 $\leq~(B-V)_0~\leq$ +0.76, where we adopt a value
of $E(B-V) =$ +0.05 for the color excess in M67 (Montgomery et al. 1993). The
resulting spectral range of the solar-type sample is $\sim$ F9 - G9, following
the $(B-V)$ color---spectral-type calibration given by Johnson (1966).
In this way, we can ensure that we are observing stars that are close
to the Sun in their photospheric properties while still retaining a relatively
large sample size.  We show in Fig. 1 a 
color-magnitude diagram of the cluster with the
survey objects indicated in order to illustrate the 
context of our investigation in the long history of observational studies of 
the M67 open cluster. 

The reduction follows standard procedures as outlined by Massey (1997) for the 
reduction of WIYN Hydra data utilizing $iraf$ routines.  In particular,  the 
CCD frames are bias-subtracted and divided by a normalized flat field frame
obtained by combining a series of dome flats.  A dispersion correction is 
applied with the aid of CuAr comparison lamp spectra acquired before and 
after each series of integrations for a given field configuration of the 
Hydra positioner.  A correction for scattered light was applied within the
iraf $dohydra$ routine.  The scattered light in the object frames is generally
at the 1--2 ADU level, or a few percent of the level of the dark H and 
K cores.  Sky subtraction was 
performed with a mean sky spectrum obtained from the individual sky fibers.
A radial velocity correction for the changing component of the Earth's rotation
(heliocentric correction) was computed and applied using the iraf routines
$rvcor$ and $dopcor$, respectively.

The results discussed herein are based on spectra obtained during the 
observing seasons from 1996--2001.  The long-term nature of the program 
enables us to search for variability that, for example, may arise from
activity cycles analogous to the $\sim$ 11-year solar cycle.  In view of the
difficulty of obtaining 
high precision measurements of the deep cores of the 
Ca II H and K resonance lines in faint stars at moderate-to-high spectral 
resolutions with moderate-aperture telescopes, we 
utilize the combined spectra obtained during this period for the purposes of 
the survey.  While this has the effect of averaging intrinsic
variations in the Ca II core strength due to chromospheric activity, it
nevertheless yields high-precision measurements of the average Ca II strength. 

\section{Calibration}

The calibration of the Hydra spectra to residual intensity follows
the empirical relations given by Hall \& Lockwood (1995) 
as functions of $(B-V)$ color. These relations give the residual intensities
at pseudocontinuum points at 3912 {\AA} and 4000 {\AA}, respectively, 
each as a function of $(B-V)$ color.  Since these calibrations were 
established using spectra at echelle resolutions, we found
it necessary to apply a small correction to the calibrations at the 
pseudocontinuum points for our somewhat lower-resolution Hydra spectra.
In particular, we degraded the calibrated solar spectrum
given in the {\it NSO Solar Flux Atlas} to the resolution of the Hydra 
spectra.  We then measured the relative intensities at the pseudocontinuum 
points at 3912 {\AA} and 4000 {\AA}, respectively, in the degraded solar
spectrum and compared the values with those computed from the relations 
cited above for the mean $B-V$ color of the Sun of +0.65 (VandenBerg \& 
Bridges 1984).  We adopted the (small) differences as constant corrections
applied to the calibrations of our stellar spectra. We display in Fig. 2 a
comparison of the NSO solar
atlas spectrum in the Ca II H\&K line region with our calibration of
the spectrum of a solar-type star observed in M67.  The correspondence of
the solar and stellar spectra confirm the self-consistent nature of our
calibration procedure.

We adopt the HK index as a measure of the chromospheric activity in the stars
in our sample.  The HK index is the sum of the relative intensities in 
1 {\AA} bandpasses centered at the H and K lines in the calibrated Hydra
spectra.  This definition is analogous to that used by 
White \& Livingston (1981) in their long-term program of 
measurement of the full disk Ca II H and K line profiles in the integrated
solar spectrum. 

Our processing of the stellar data includes the use of a high quality lunar 
spectrum, obtained with
the Hydra bench spectrograph in our configuration, as a template.  The stellar
spectra are shifted in wavelength until the lunar and each stellar
spectrum are aligned in the Ca II H \& K region.  We define the central
wavelengths for the H and K lines by the location of the minima of the
H \& K features in the lunar template.  In this way, the locations in
wavelength of the bandpasses for the computation of the HK index can be      
established in an objective manner and applied uniformly to all the
stellar spectra.  Likewise, the locations of the psuedocontinuum 
points are identified as the maximum values in the observed lunar spectrum to
within about one-half a resolution element of 3912 {\AA} amd 4000 {\AA}, 
respectively.  The wavelengths of these locations are used to 
find the corresponding psuedocontinuum points in the stellar spectra for
the purposes of normalization and calibration.  The sum of the relative 
intensities in the H and K line cores are then computed for the calibrated
spectra 
with the aid of a spline fit to the pixel values within the 1 {\AA} bandpasses,
including the interpolated values at the locations of the bandpass boundaries 
at $\pm$0.5 {\AA} from the fixed central wavelength values.

In a similar manner, the random errors due to photon noise 
are estimated from the uncalibrated spectra (in ADUs)
and taking into account the gain of the CCD.  That is, the noise estimate is
based on the sum of the 
estimated photon counts inferred from a combination of the CCD gain and 
a smooth spline fit
to the pixel values within the respective 1 {\AA} bandpasses for the
H and K lines (including the interpolated bandpass boundaries).
In addition, a systematic overestimate
of the H \& K fluxes is present in our results due 
to ``smearing" of any core emission as a result of the use of only 
moderate spectral resolution.  In order to estimate the magnitude of this 
effect, we degraded a high resolution solar spectrum to the 
0.8 {\AA} resolution of the Hydra bench spectrograph. We found that the 
measured HK index increased 
by 4.7\% relative to the value we measured for the undegraded spectrum.
We therefore take this factor into account when comparing the solar data to
our M67 data in the following discussion.  In particular, we adjust the
measured solar HK index values upward by 4.7\% when comparing the solar
data to our M67 data so that both data-sets are on the same scale. 

\section{Results}

 We show in Table 1 the results for our survey of chromospheric Ca II H+K 
core strengths in our sample of solar-type stars in M67.  The mean
HK index and error in the mean (inferred from the quadratic sum of the errors 
in the individual spectra) are listed for each program star by 
Sanders number.  
We also 
give the de-reddened $B-V$ color, the apparent visual magnitude (from 
Montgomery et al. 1993), and
the membership probabilities (as described above).  The final column  
indicates the binary status of the object according to information
kindly provided by R. Mathieu in advance of publication.  No entry in this
column means that no information is available.
We note, parenthetically, that we exclude S1112 from the following analysis
and discussion.  This object, which exhibits strong H and K line emission
in our spectra, is a detected X-ray source and a suspected RS CVn binary 
system (see note to Table 1).
In addition, we utilized unassigned fiber cables in the Hydra instrument to 
obtain spectra of M67 dwarf stars that are just outside 
the intrinsic $B-V$ color range we adopt as our criterion for
solar-type.  We provide the results for these stars in
Table 1 for their potential utilization in other investigations.
However, we do not include these objects in the analysis and
discussion given herein. 

 The results for the remaining 60 stars in Table 1 that meet our
photometric criteria given in $\S$2 for solar-type are encapsulated 
in the histogram in Fig. 3, which 
shows the distribution of the values of the HK index.  In order to place 
the stellar observations in context, we also display 
the distribution of the HK index for the solar cycle as measured from
1976 to 2004 (Livingston 2005, private communication) but
adjusted upward by 4.7\% in order to take into account the substantial
difference in spectral resolution between the two data-sets, as discussed
in $\S$3. Inspection of Fig. 3
reveals that the distribution of HK index for the solar-type 
stars in M67 is broader than the range of the contemporary solar cycle.  
The stellar distribution also appears unimodal and slightly skewed with 
a tail toward
higher activity stars.  We note that the solar cycle distribution is slightly
skewed toward higher activity levels, as is the distribution of 
Mt. Wilson S-index values for the sample of field solar-type stars 
investigated by Hall \& Lockwood (2004; see their Fig. 2).  This could arise 
if the decay from maximum in the individual stellar cycles is longer than
the rise time from minimum. The shape of the $\sim$ 11-year solar cycle 
exhibits this same characteristic.

 A comparison of the stellar with the solar distribution shows that about 17\%
of the solar-type stars in the M67 sample exhibit values of the HK index less 
than
solar minimum; $\sim$ 7\% are in excess of solar maximum while approximately
77\% exhibit values of the HK index within the range of the 
solar cycle.  Thus, approximately 24\% of the M67 sun-like stars are 
characterized by a 
level of chromospheric activity that is outside the range of the contemporary 
solar cycle while significant overlap with the range of the solar cycle is 
present in 80\% of the stellar sample. If we exclude the slight extension of
the solar data to the unusually high HK index of approximately 230 m{\AA} 
and instead adopt a more representative maximum value 
of HK $\approx$ 225 m{\AA} (from inspection of Fig. 3), 
then the fraction of stars that exceeds the solar activity maximum 
becomes approximately 12\%.  The fraction that is within the range of 
the solar cycle decreases to about 72\%.


 We examine in Fig. 4 the color dependence of the HK index as a function of
intrinsic $B-V$ color.  We only note that the scatter in HK index appears
to increase toward the lower mass stars in the sample.  This kind of 
increasing dispersion in a magnetic activity diagnostic toward  lower mass
is a feature of late-type dwarf stars in clusters, at least as seen in 
H$\alpha$ (e.g, Stauffer et al. 1991).  In order to gain further insight 
on the contributions to the total M67 distribution in Fig. 
3 by color bins, we display a sequence of histograms 
in Fig. 5 delineated according to subsets in $B-V$ intrinsic 
color.  The histograms in Fig. 5 exhibit considerable overlap with 
the solar cycle distribution.  However, departures from the cycle
distribution toward both higher activity levels exceeding solar maximum and 
levels below solar minimum appear more prevalent toward later spectral
types.   

 It has long been recognized that binary membership can enhance and
prolong magnetic field-related activity.  We show in Fig. 6 the distribution of
the HK index in the M67 sample with the contribution by known binaries
separately indicated.  The HK index distribution for the binaries
exhibits significant overlap with the distribution of the M67
solar-type stars that are presumably single (or for which no radial
velocity data are yet available).  The binary star distribution in HK strength 
appears to be well segregated between relatively low and high values.
However, we hesitate to attribute any physical significance to this
apparent gap given the small
sample size.   Finally, we give in Fig. 7 a plot of the HK index
versus period for the binaries in our sample with known orbital periods, kindly
provided in advance of publication by Mathieu (2005, private communication).  
In contrast to the active RS CVn binary systems in 
M67 (Belloni, Verbunt \& Mathieu 1998; van den Berg et al. 2004), no 
correlation between chromospheric Ca II core strength and orbital period 
is apparent in the binary stars in our solar-type sample.

\section{Discussion}

 The broader distribution in Ca II H+K core strengths in the M67 solar-type 
stars, compared to that of the Sun during the contemporary epoch, suggests 
that the potential excursion in the amplitude of the
solar cycle is greater than what we have seen so far.  
The stars with values of the HK index noticeably smaller than
that of solar minimum may be
in a prolonged state of relative quiescence analogous to the
Maunder Minimum of the Sun during
1645-1715 C.E. when visible manifestations of solar
activity, e.g., sunspot number, substantially decreased
for a period much longer than the sunspot cycle. That
persistent period of reduced solar activity occurred
during the extreme phases (ca. mid-16th through 17th century)
of a climatic anomaly characterized by a general cooling and
referred to as the Little Ice Age.  
That anomaly, like the Medieval Warm Period which 
preceded it, is climatically,
spatially and temporally complex, with evidence of regional variations 
in conditions indicating a sensitivity to a variety of forcings,
including that due to a variable Sun (Rind 2002).
The plausible assumption of a multi-decadal period
of decreased total solar irradiance accompanying the observationally
established low levels of solar magnetic activity has been 
postulated as a contributing factor to the
sharply deteriorated conditions of the 17th-century
(see, e.g., Eddy 1976; Foukal \& Lean 1990; Hoyt \& Schatten 1997;
Soon \& Yaskell 2004).  However, quantitative estimates of the presumed
decrease in the solar irradiance during this time are uncertain;
recent work (Wang, Lean \& Sheeley 2005; Lean et al. 2005) concludes 
that the secular increase in the total
irradiance from the Maunder Minimum to the current cycle minima is less
than the initial estimates that appeared in the literature beginning
over a decade ago. 

The $\sim$ 17\% fraction of M67 
sun-like stars in our sample that is at this relatively quiescent level 
would represent the frequency at which the Sun enters a Maunder
Minimum. 
This frequency is lower than the 
estimated frequency that the Sun exhibits especially low levels
of magnetic field-related activity.  In particular, Damon (1977)
inferred from the proxy terrestrial record of solar magnetic
activity based on the $^{14}$C radioisotope that the Sun has spent
about one-third of the time in the past several millennia in
magnetic activity minima.  Our results for the fraction of M67 solar-type 
stars that exhibit especially quiescent levels of activity is also lower
than the $\sim$ one-third frequency seen in samples consisting of field
solar-type stars (Baliunas \& Jastrow 1990; Hall \& Lockwood 2004).
However, it is higher than the fraction Wright (2004) finds when subgiants 
are removed from the Mt. Wilson field star sample.

 Our M67 sample of solar-type stars, which is presumably homogeneous in age and
metallicity, does not exhibit the bimodal distribution that 
appeared in the heterogeneous sample of field stars studied by
Baliunas \& Jastrow (1990).  From a purely qualitative perspective,
our results imply that exceptional quiescence, such as that during a
Maunder Minimum-like episode, is simply a low-amplitude extension of
the solar dynamo rather than a separate dynamo mode.  In this regard,
Ribes \& Nismes-Ribes (1993) concluded from their analysis of
historical observations that the solar cycle persisted through the
Maunder Minimum, though at exceptionally low amplitudes in terms of
sunspot number.  Likewise, the occurrence of activity levels
that exceed contemporary solar maximum values could be a
manifestation of the excursion to enhanced levels of activity in a single 
mode of dynamo operation.  Our results ($\S$4) would suggest that 
exceptionally high levels of activity can occur at a frequency of
$\sim$ 7\% - 12\% of the time.

 We present in Fig. 8 another depiction of the comparison
between the solar cycle and the M67 solar-type star distributions, 
respectively, in the form of cumulative probability functions of the HK index.  
The curve for
the M67 sample reflects the broader distribution in HK index compared
to the solar cycle.  Inspection of Fig. 8 readily reveals that the
median values for the M67 and the modern solar cycle distributions,
respectively, are nearly identical.
Finally, a Spearman rank correlation test applied to the two distributions in
Fig. 8 yields a moderately high correlation at $\rho$ = 0.881.
However, if we apply this test after omitting the
``supersolar" stars with HK $>$ 300 m{\AA} then we find an extremely
high correlation at $\rho$ = 0.978.  This result, while not conclusive,
is consistent with an activity-cycle origin, analogous to the solar cycle,
for the observed HK index distribution for the solar-type stars in M67.  


 The HK index distribution in M67 according to color, as illustrated 
in Figs. 4-5, 
merits further consideration.  The distribution in Fig. 5(a) appears skewed 
with a tail toward higher activity stars with HK $\gtrsim$ 200 m{\AA}.  A
comparison with Fig. 4 reveals that the peak in the distribution consists
mainly of bluer stars
with intrinsic $B-V$ colors $\lesssim$ 0.60.  These objects may
indeed be relatively more chromospherically active in this color
range.  However, it is important to recognize that the
non-chromospheric, radiative equilibrium contribution to the core flux
in the Ca II H \& K lines will be relatively higher in hotter stars
and decline toward cooler objects (Linsky et al. 1979; Noyes et al. 1984).  
A reliable calibration of the
radiative equilibrium flux for the 1 {\AA} bandpasses that
define the HK index has not yet been done.

 The distribution in Fig. 5(b) is especially interesting because this subsample
is the most photometrically similar to the Sun.  In particular, the range
of colors that has been quoted in the literature for the Sun is 0.63
$\leq~B-V~\leq$ 0.67 (VandenBerg \& Bridges 1984, see their Table 2).
Hence, the 21 M67 sun-like stars we observed in this color range are 
photometric counterparts
of the Sun and thus can be considered true
solar analogs, following the definition of a solar analog as given by 
Cayrel de Strobel \&
Bentolilla (1989; also see Cayrel de Strobel 1996).  In the case of these 
M67 solar analogs, 2 stars or 
10\% are in a state of enhanced activity with HK indices in excess of solar 
maximum values while 4 stars (about 19\%) exhibit values of HK that are below 
that of the contemporary solar minimum HK index; 71\%, or 15 stars, have 
an average HK index that is within the range of the modern solar cycle.
Thus, 29\% of the solar analogs are characterized by levels of chromospheric
activity that are outside the range of the modern solar cycle.
In brief summary, a significant fraction of this homogeneous sample of 
solar analogs exhibits activity that is, in amplitude, consistent with the 
solar cycle as seen in the chromospheric Ca II resonance lines.
The fraction that is outside of the cycle range can be indicative of the 
frequency of excursions in the solar cycle 
to either enhanced levels of activity or unusually quiescent episodes 
(e.g., Maunder minima).

 Those stars that are slightly cooler and less massive than the Sun in 
the color range of 0.68 $\leq~(B-V)_0~\leq$ 0.72 also show substantial 
overlap with the solar cycle with 6/11 stars in the cycle range; 3-4 stars 
are more active than solar maximum and 1-2 objects have values of the HK index 
less than
the contemporary solar minimum.  
Similarly, the coolest stars observed in our sample with intrinsic $B-V$
colors in the range of 0.73 -- 0.76 include 2 stars more quiescent
than solar minimum, no objects more active than solar maximum and 4 stars
in the range of the solar cycle.   In brief summary, the solar-type stars
in this approximately solar-age and solar-metallicity cluster
exhibit levels of chromospheric activity, as represented by our HK index,
that are substantially within the range of the solar cycle.  Excursions
outside of the solar cycle range of activity occur at a frequency of about
29\% for the entire solar-type sample, and at frequencies broadly
in the range of $\sim$ 22\%--45\% for the subsets of the sample given
in Fig. 5.

\subsection{Implications for Brightness Changes}

 Given the correlation between variations in chromospheric emission and
changes in brightness or irradiance, as in the case of the Sun, it is 
of interest to examine the implications of the range of chromospheric 
activity we see in M67 for the possible range of brightness amplitude 
variations that
may occur.  In their long-term study of the photometric variability of 
field solar-type stars, Lockwood, Skiff \& Radick (1997; their Fig. 16) 
present the mean 
amplitude of variation in photometric brightness as 
a function of log $R^{\prime}_{HK}$, i.e., the
net chromospheric radiative flux in the H + K lines, normalized by the
stellar bolometric flux (= $\sigma T_{eff}^4$).

In order to estimate the potential range in amplitude of brightness changes in 
our sample of M67 sun-like stars, 
we must first estimate
the net chromospheric radiative loss rates in the H \& K lines from the
HK index and the stellar effective temperature.  We utilize the calibrations
given by Hall \& Lockwood (1996) as a function of $B-V$ color in order to
estimate the effective temperature and to obtain the total surface flux in
the H \& K cores from the HK index.  An empirical correction for the radiative
equilibrium (photospheric) contribution to the H and K core flux 
as a function of $B-V$ color is given by Noyes et al. (1984).
This approach yields an estimate of $R^{\prime}_{HK}$. 

Using our estimates of this parameter from above, we find a range in 
log $R^{\prime}_{HK}$ of $\sim$ -4.35 to -5.1.  From the results of 
Lockwood et al. (1997) this range in log $R^{\prime}_{HK}$ corresponds to
brightness variations 
extending from sun-like ($\sim$ 0.1\%) to nearly 3\%, or considerably in 
excess of that of the contemporary Sun.  Whether long-term brightness
variability of
these amplitudes actually occurs in the M67 solar-type stars will require
verification through high-precision photometric monitoring. 

\subsection{Variability}

 The acquisition of the data for this program over several observing seasons
enables us to conduct a preliminary investigation of the nature of long-term 
variability in the HK index in the solar-type stars in M67.  We display in
Fig. 9(a)-(n) the variation of the HK index by season in those program objects
for which we have measurements in at least two observing seasons.  The panels 
are ordered by decreasing $B-V$ color and span the 1996-2002 
seasons.  We also include in Fig. 9, for comparative purposes, the annual 
mean values of the HK index for the Sun during this same period.  

 Inspection of Fig. 9 reveals the presence of variability in the HK index that 
exceeds the formal errors in practically all the program objects.  The time
series are not sufficiently long to enable us to conclusively identify the
presence of magnetic activity cycles, analogous to the solar cycle, and to 
measure actual cycle periods, though some cases are suggestive
of a cycle-like variation.  It does appear from the results displayed in 
Fig. 9 that if activity cycles are present then their periods are greater 
than $\sim$ 6 years.  

We include in Fig. 10 further examples of comparable 
portions (in time) of the solar
cycle variation based on mean annual HK index values for comparison with 
the panels in Fig. 9.  In general, the
variability seen in the M67 solar-type stars appears to be characterized by
higher amplitudes.  We summarize the seasonal variability in HK index in 
Table 2 for our program objects.  The rms variation in HK index among the 
M67 solar-type stars ranges from 4.60 m{\AA} to 92.1 m{\AA} with a mean 
rms of 19.9 m{\AA}.  We also include in Table 2 the rms variations for the
Sun for the examples given in Fig. 10 as well as for the period from 
1976-2004.  In the case of the latter, the rms deviation of the annual
mean HK index for the Sun is 7.90 m{\AA}, which is within the range seen 
for the M67 solar-type stars though significantly lower than the mean rms
for the sample.  Thus, we find that, while there are 
seasonal variations comparable to that seen in the Sun in some M67 
solar-type stars, variability in the HK index substantially in excess to 
that of the Sun dominates our results.  We do not see a clear correlation 
between root mean square deviation and the HK index values in 
the entire data-set (Fig. 11).  However, inspection of Fig. 11 suggests a 
possible correlation of increasing rms deviation with HK index for 
values of rms  $\gtrsim$ 20 m{\AA}.   A more precise study of
variability will require a long-term program, ideally with spectra obtained
at higher spectral resolution in order to accurately assess
the nature of the variability in the H and K line cores.

 We construct in Fig. 12 the histogram of all the 
HK index values measured seasonally for the solar-type stars.  
Not surprisingly, the distribution is broader than that in Fig. 3, which
is based on the HK index for each star averaged over several seasons.
We find in Fig. 12 that 20\% of the HK index measurements are 
less than the modern
solar minimum value while Ca II core strengths in excess of solar 
maximum occur at a frequency of 11\%.  Thus, about 31\% of the HK index
measurements are outside the range of the solar cycle in our sample of M67 
solar-type stars when examined on a seasonal basis.

\subsection{Chromospheric Activity and Stellar Age}

 The calibration of the empirical correlation between chromospheric 
activity and stellar age (Skumanich 1972) can yield an additional tool 
for the determination of the ages of field stars.
It is therefore important to investigate quantitatively the magnitude 
of the impact of 
intrinsic variations in chromospheric activity, such as magnetic cycle 
variability, on the calibration of this relation.

 Relations between the net radiative chromospheric flux parameter in the 
Ca II H \& K lines, $R^{\prime}_{HK}$, 
and age have been 
given by Soderblom et al. (1991) and subsequently refined and extended 
to younger ages $\lesssim$ 1 Gyr by Donahue (1993; also see Donahue 1998).  
We adopt this age-activity calibration to examine the {\it apparent} age spread
implied by the dispersion in chromospheric activity in both M67 and the
solar cycle.  In the case of the solar cycle, we utilize the HK index values
based on the actual measurements from the high resolution solar spectra in
order to give a more realistic estimate of the solar chromospheric age. 

Inserting our estimates
of $R^{\prime}_{HK}$ from $\S$5.1 into the age-activity calibration given 
by Donahue (1998) yields the results in Fig. 13, shown both for the solar-type 
stars in M67 and the contemporary solar cycle.  The broad extent of the 
apparent age range due to 
the cycle variation in Ca II in the Sun is $\sim$ 2.5 Gyr to about 6 Gyr.  
The mean chromospheric age of the Sun is approximately 4.26 Gyr. 
We note, parenthetically, that Baliunas et al. (1998) similarly estimated 
the apparent (chromospheric) age, or error in estimating the age of the 
Sun over the last four hundred years if chromospheric activity were sampled at 
a random phase of long-term variability, namely, 3 to 8 Gyr at the extremes 
with presumably a narrower range in apparent age for most of the period.
The age range among the M67 solar-type stars implied by the range of observed 
chromospheric activity is even broader, extending from less than
1 Gyr to about 7.5 Gyr.  The mean chromospheric age inferred from our sample
of solar-type stars is approximately 3.75 Gyr while the median age for the
M67 distribution is about 3.33 Gyr.  Taking into account the 
possibility of a systematic bias in our stellar HK index values ($\S3$) 
yields a 
mean chromospheric age of 4.32 Gyr and a median of 3.83 Gyr.  In summary, the 
mean chromospheric age for the solar-type members 
of M67 considered herein is in the range of roughly 3.8 - 4.3 Gyr while
that for the Sun is approximately 4.3 Gyr.  Clearly,
age determinations based on chromospheric emission can vary considerably, 
depending for each star on the phase of its magnetic activity 
cycle at the time of observation and, in the case of cluster members that
are presumably homogeneous in age, on the dispersion in rotational 
velocities that may be present.

\subsection{The Range of Chromospheric Activity in the M67 Solar-Type Stars}

 A surprising result of our investigation is the occurrence of solar-type
stars that exhibit levels of chromospheric activity in excess of that seen
at the maximum of the contemporary solar-activity cycle.  An extreme example 
is illustrated in Fig. 14 where Ca II H and K line core emission is 
clearly seen in this spectrum of S1452.  This object is not known to
be a binary at a precision of $\pm$2 km s$^{-1}$ in the possible variation of 
its radial velocity (R. D. Mathieu, private communication).

 An interpretation of our results is that the origin of the dispersion in
chromospheric activity among the solar-type stars in M67 is due to the 
range of amplitudes in their cycle properties.  But we also know, in 
general, that 
magnetic field-related chromospheric activity increases with increasing 
equatorial rotational velocity in late-type stars.  Thus, those M67 
solar-type stars with relatively enhanced levels of activity, compared 
to the Sun, may be rotating more rapidly, thereby giving rise to stronger
dynamo action.  If the chromospheric Ca II H and K strength is proportional
to the rotational velocity (following Skumanich 1972) then this would 
suggest for the high-activity M67 stars in Fig. 3--with roughly twice the
HK index as the mean Sun--that their rotational velocities are in the 
3-4 km s$^{-1}$ range, or slightly more.

 Obviously, this issue can only be addressed through measurements of (a) 
rotation period, via ultra-high precision photometric observations of spot
modulation, (b) through spectrophotometric observations of the rotational 
modulation of the H and K lines, or (c) by spectroscopic measurements of the
projected rotation velocity.  If the results ultimately reveal a significant
spread in rotation rates then this would argue against activity cycles as 
the principal origin of the distribution of activity in Fig. 3.  Moreover, 
it would raise the important question of why the angular momentum histories 
of the Sun and the M67 solar counterparts differ.  If, however, the M67
sun-like stars exhibit rotational velocities that are more solar-like 
($\sim$ 2-3 km s$^{-1}$) then this would suggest that activity cycles--similar
to the modern solar cycle but characterized by significant differences in
amplitude--are the origin of the dispersion in chromospheric activity we 
see in Fig. 3.  However, this would then raise the question of why the 
contemporary Sun is in a relatively subdued state of activity compared
to other sun-like stars of solar age and solar metallicity. 

\section{Summary and Future Directions}

 We summarize in the following our principal findings and discuss our 
future directions of research based on these results.

 We find that the HK index distribution for the solar-type stars in M67 is 
broader than that of the contemporary solar cycle though with significant 
overlap. More specifically, $\sim$ 72\% - 87\% of the 60 solar-type stars in
our sample exhibit mean HK index measures, averaged over several seasons of
observation, that are within the range of the modern solar cycle.
The 21 solar analogs we observed in our M67 sample, i.e., stars that 
are photometric analogs of the 
Sun, exhibit a similar correspondence to the solar cycle distribution
in HK index, with 71\% of these objects characterized by an average HK index
within the range of the contemporary solar cycle.  The results for the 
seasonal distribution
of HK index are similar in terms of the frequency of values that are
within and outside the range of the modern solar cyle.  
We interpret the fraction
of the M67 sample that is outside the cycle range as indicative of the 
frequency and nature of excursions in the solar cycle to either enhanced 
levels of activity or to exceptionally quiescent episodes reminiscent of a
Maunder Minimum.  

 The HK index distribution in the M67 solar-type stars is not bimodal, 
in contrast to the earlier study by Baliunas \& Jastrow (1990).  We suggest
that our result is a manifestation of a single-mode of dynamo operation with
occasional extensions in amplitude to either enhanced levels of 
activity or unusually quiescent episodes.

 The inferred range in the possible amplitudes of long-term 
brightness changes extends from sun-like at $\sim$ 0.1\% to as high as 
3\%, or similar to the amplitude of photometric variability in 
Hyades-age stars (Radick et al. 1995).

 Seasonal variability is observed in the program objects with a 
mean rms deviation in the HK index that is more than a factor of 2 higher
than what is observed in the contemporary solar cycle.  In general, the 
amplitude of variation in chromospheric activity is higher in the M67 
solar-type stars than in the Sun.  We have not yet identified conclusively 
the occurrence of cycle variability analogous to that of the solar cycle  
though some program objects exhibit long-term trends suggestive of cycle-like 
variability.  In any event, if activity cycles are present then our data
indicate that the cycle periods are greater than $\sim$ 6 years.  

 The chromospheric age range in the M67 sample extends
from less than 1 Gyr to about 7.5 Gyr.  This result is indicative
of the magnitude of the impact of intrinsic chromospheric variations on age
determinations based on the observed
levels of activity in a stellar sample that is ostensibly homogeneous in age.
We estimate a mean chromospheric age for M67 in the range of
approximately 3.8--4.3 Gyr.

 Our intriguing results pose crucial questions. In particular, is the 
HK index distribution of chromospheric activity in the sun-like stars in 
M67 really the result of the long-term modulation of activity by cycles
analogous to the solar cycle, or are the relative amplitudes of the
cycles actually similar with the differences due only to differences in
the mean level of activity, perhaps resulting from a dispersion in 
rotational velocities?  Even more fundamentally, is the range and variability
of chromospheric activity as seen in the cores of the H and K lines arising
from sunlike cycles at all?  These critical issues can only be addressed
through regular observations of the ``suns of M67" over a period of several
years.  Ideally, such a program should be conducted at echelle resolutions
in order to more precisely measure the line strength and line profile 
variations in the cores between the K$_1$ (H$_1$) minima.  In addition, 
high-precision differential photometry obtained in parallel with
this kind of program would yield a direct determination of the nature of
the joint variation of brightness and magnetic field-related activity.

 The presence of very active solar-type stars in this solar-age cluster, 
that are also apparently single, is perplexing.  We are in the process of
attempting to obtain projected rotational velocity measurements 
in order to determine if it is rapid rotation that is the origin of the
relatively enhanced activity in these objects.  If this is verified to be
the case then it would invite further investigation, both theoretical and
observational, of the angular momentum history of M67 in contrast to that
of the Sun and other solar-age G dwarfs.  However, if the velocities are
determined to be solar-like (i.e., $\sim$ 2 km s$^{-1}$) then we can 
conclude that excursions to enhanced activity levels that are significantly in 
excess of contemporary solar maximum values occur in sun-like 
stars and perhaps in the Sun itself.

\acknowledgments{We are grateful to the NOAO Galactic TAC for its
support of this long-term program.  We are especially grateful to
Dr. Diane Harmer for her assistance with the acquisition of the data for this
investigation during the course of the WIYN Queue program and thereafter.  We 
also acknowledge with appreciation the contributions by Mr. Daryl Willmarth, 
Dr. Paul Smith and Dr. Abi Saha who each obtained data for our 
investigation during the WIYN Queue program.  
We are particularly grateful to Dr. Bill Livingston 
for sharing with us his measurements of the solar H and K index from his
multi-decadal program of integrated sunlight observations at the McMath-Pierce
Solar telescope.   We acknowledge interesting
discussions with Dr. Claus Fr\"{o}hlich and Dr. Peter Foukal during the
course of this work. The authors thank
Dr. Bob Mathieu for sharing in advance of publication his findings 
concerning binarity among solar-type stars in M67.  We thank the referee, Dr. 
Doug Duncan, for his thoughtful review of the original manuscript.
Finally, SLB gratefully
acknowledges support by grants from the Richard C. Lounsbery Foundation and 
JPL grant P6201-7-05.}


\newpage

\begin{deluxetable}{l c c c l c c l l}
\tablenum{1}
\tablecolumns{9}
\tablecaption{Survey Results}
\tablehead{\colhead{Sanders} & \colhead{$(B-V)_o$} & \colhead{HK} & \colhead{Error} & \colhead{V} & \colhead{P$_\mu$} & \colhead{P$_{{\mu},r}$} & \colhead{binarity\tablenotemark{a}}\\
 \colhead{No.} & \colhead{ } & \colhead{(m\AA)} & \colhead{(m\AA)} & \colhead{ } & \colhead{ } & \colhead{ } & \colhead{ }}
\startdata
603	&	0.56	&	208	&	9.99	&	14.05	&	97	&	96	&	--\\
621	&	0.60	&	200	&	8.3	&	14.44	&	87	&	86	&	--\\
724	&	0.63	&	200	&	10.8	&	14.54	&	93	&	85	&	--\\
746	&	0.66	&	209	&	8.58	&	14.38	&	94	&	96	&	--\\
747	&	0.65	&	354	&	10.9	&	14.052	&	90	&	95	&	--\\
748	&	0.78	&	264	&	12.3	&	14.632	&	89	&	94	&	y\\
753	&	0.58	&	208	&	7.49	&	14.693	&	93	&	96	&	**\\
770	&	0.63	&	195	&	7.43	&	14.636	&	88	&	96	&	**\\
777	&	0.63	&	208	&	8.7	&	14.52	&	93	&	96	&	**\\
779	&	0.68	&	190	&	8.34	&	14.43	&	85	&	94	&	**\\
785	&	0.65	&	221	&	8.24	&	14.823	&	88	&	95	&	**\\
789	&	0.61	&	204	&	5.99	&	14.054	&	85	&	95	&	--\\
801	&	0.67	&	182	&	6.71	&	15.094	&	87	&	95	&	**\\
802	&	0.67	&	193	&	8.08	&	14.792	&	86	&	93	&	**\\
829	&	0.54	&	172	&	7.49	&	14.37	&	91	&	87	&	--\\
937	&	0.59	&	194	&	8.52	&	14.12	&	95	&	88	&	--\\
942	&	0.58	&	206	&	11.5	&	14.475	&	90	&	87	&	--\\
943	&	0.71	&	166	&	18	&	14.96	&	85	&	82	&	--\\
945	&	0.62	&	205	&	7.65	&	14.528	&	93	&	92	&	**\\
951	&	0.67	&	190	&	7.09	&	14.704	&	87	&	87	&	y\\
958	&	0.62	&	206	&	8.52	&	14.45	&	96	&	97	&	--\\
963	&	0.66	&	168	&	11.3	&	14.513	&	91	&	96	&	P=90.28\\
965	&	0.71	&	222	&	11.6	&	14.695	&	80	&	90	&	**\\
969	&	0.62	&	203	&	9.07	&	14.177	&	87	&	95	&	**\\
981	&	0.66	&	177	&	14.9	&	14.16	&	96	&	99	&	P=56.0\\
982	&	0.56	&	231	&	6.99	&	14.122	&	94	&	99	&	P=373\\
991	&	0.63	&	179	&	8.67	&	14.564	&	91	&	97	&	**\\
1004	&	0.71	&	214	&	10.5	&	14.928	&	89	&	98	&	2\\
1012	&	0.69	&	228	&	10.3	&	14.516	&	88	&	98	&	P=641\\
1014	&	0.66	&	248	&	13.8	&	14.183	&	86	&	98	&	P=16.2\\
1033	&	0.56	&	209	&	8.69	&	14.164	&	97	&	99	&	**\\
1041	&	0.68	&	193	&	9.51	&	14.718	&	83	&	97	&	2\\
1048	&	0.64	&	188	&	8.95	&	14.411	&	92	&	99	&	--\\
1050\tablenotemark{d}	&	0.61	&	390	&	19.7	&	14.292	&	84	&	97	&	y\\
1057	&	0.63	&	210	&	7.03	&	14.303	&	81	&	96	&	**\\
1064	&	0.61	&	178	&	11.2	&	14.04	&	94	&	99	&	P=575\\
1065	&	0.75	&	222	&	8.99	&	14.645	&	77	&	93	&	P=150\\
1068	&	0.70	&	211	&	7.84	&	15.028	&	83	&	95	&	2\\
1078	&	0.61	&	190	&	7.13	&	14.174	&	97	&	99	&	**\\
1087	&	0.59	&	198	&	8.13	&	14.16	&	84	&	94	&	--\\
1089	&	0.62	&	173	&	7.31	&	14.199	&	91	&	96	&	--\\
1093	&	0.59	&	207	&	8.94	&	14.134	&	83	&	91	&	--\\
1095	&	0.61	&	183	&	6.74	&	14.59	&	88	&	90	&	**\\
1096	&	0.62	&	192	&	7.23	&	14.49	&	94	&	95	&	--\\
1106	&	0.67	&	168	&	9.72	&	14.78	&	91	&	90	&	--\\
1107	&	0.55	&	241	&	7.5	&	14.13	&	96	&	95	&	--\\
1112\tablenotemark{b}	&	0.69	&	729	&	18.5	&	15.05	&	92	&	86	&	--\\
1203\tablenotemark{c}	&	0.67	&	218	&	10.4	&	14.413	&	97	&	97	&	--\\
1208	&	0.79	&	229	&	18.9	&	14.6	&	92	&	95	&	P=19.9\\
1212	&	0.73	&	218	&	13.1	&	15.32	&	83	&	86	&	1\\
1213	&	0.55	&	189	&	8.64	&	14.114	&	98	&	99	&	--\\
1218	&	0.63	&	194	&	9.7	&	14.59	&	97	&	99	&	**\\
1246	&	0.64	&	187	&	9.65	&	14.623	&	82	&	95	&	**\\
1247	&	0.57	&	222	&	5.59	&	14.044	&	95	&	99	&	P=69.8\\
1248	&	0.57	&	202	&	8.82	&	14.22	&	96	&	99	&	**\\
1249	&	0.73	&	203	&	9.31	&	14.314	&	92	&	97	&	--\\
1251	&	0.70	&	231	&	10.4	&	14.79	&	81	&	90	&	**\\
1252	&	0.59	&	201	&	8.14	&	14.067	&	97	&	99	&	**\\
1255	&	0.62	&	211	&	6.75	&	14.486	&	94	&	99	&	**\\
1258	&	0.61	&	183	&	8.38	&	14.54	&	92	&	97	&	**\\
1260	&	0.58	&	202	&	8.2	&	14.191	&	97	&	99	&	**\\
1269	&	0.71	&	183	&	8.18	&	14.934	&	81	&	93	&	**\\
1278	&	0.73	&	181	&	11.9	&	14.401	&	91	&	98	&	SB2, long\\
1289	&	0.71	&	176	&	6.61	&	14.901	&	94	&	99	&	**\\
1307	&	0.76	&	172	&	13.2	&	15.167	&	80	&	91	&	1\\
1318	&	0.51	&	217	&	7.05	&	14	&	90	&	92	&	--\\
1330	&	0.55	&	238	&	8.57	&	14.04	&	82	&	77	&	--\\
1341	&	0.62	&	233	&	8.65	&	14.68	&	94	&	80	&	--\\
1406	&	0.55	&	201	&	7.75	&	14.02	&	96	&	87	&	--\\
1420	&	0.56	&	205	&	8.29	&	14.15	&	95	&	92	&	--\\
1426	&	0.57	&	197	&	8.6	&	14.25	&	91	&	82	&	--\\
1446	&	0.57	&	207	&	7.42	&	14.036	&	74	&	72	&	--\\
1449	&	0.61	&	198	&	11.7	&	14.381	&	96	&	97	&	--\\
1452	&	0.62	&	414	&	11.1	&	14.578	&	86	&	88	&	**\\
1462	&	0.63	&	193	&	6.98	&	14.37	&	97	&	98	&	--\\
1473	&	0.73	&	158	&	11.5	&	15.055	&	83	&	86	&	2\\
1477	&	0.67	&	181	&	9.12	&	14.593	&	91	&	93	&	**
\enddata
\tablenotetext{a}{R. D. Mathieu, private communication. A dash indicates that
no radial velocity information is available; quoted periods {\rm P} are in 
units of days; ``SB2, long" denotes a long-period, double-line spectroscopic 
binary; 1 = one observation 
at the cluster mean radial velocity; 2 = rms $<$ 2 km s$^{-1}$, with 
observations in two years; y = binary, no orbital elements; ** = rms $<$ 
2 km s$^{-1}$}
\tablenotetext{b}{$ROSAT$ X-ray source and possible RS CVn; Belloni et al. 
(1998).  Object not included in analysis presented herein.}
\tablenotetext{c}{Star on binary sequence. Possible X-ray source; van den 
Berg et al. (2004)}
\tablenotetext{d}{$Chandra$ X-ray source; van den Berg et al. (2004)}
\end{deluxetable}
%
%
%
\begin{deluxetable}{l c c}
\tablenum{2}
\tablecolumns{3}
\tablewidth{0pt}
\tablecaption{Chromospheric Variability of M67 Solar-Type Stars}
\tablehead{\colhead{Sanders} & \colhead{$(B-V)_o$} & \colhead{RMS}\tablenotemark{a}\\
 \colhead{No.} & \colhead{ } & \colhead{(m\AA)}}
\startdata
621	&	0.60	&	12.7\\
724	&	0.63	&	22.4\\
746	&	0.66	&	9.55\\
747	&	0.65	&	22.0\\
770	&	0.63	&	13.0\\
777	&	0.63	&	16.0\\
779	&	0.68	&	8.44\\
785	&	0.65	&	14.0\\
789	&	0.61	&	4.6\\
801	&	0.67	&	15.0\\
802	&	0.67	&	7.66\\
937	&	0.59	&	8.16\\
942	&	0.58	&	16.4\\
945	&	0.62	&	21.6\\
958	&	0.62	&	14.3\\
963	&	0.66	&	15.0\\
965	&	0.71	&	13.3\\
969	&	0.62	&	19.4\\
981	&	0.66	&	5.00\\
991	&	0.63	&	14.7\\
1004	&	0.71	&	30.3\\
1014	&	0.66	&	40.0\\
1041	&	0.68	&	31.3\\
1048	&	0.64	&	16.3\\
1057	&	0.62	&	11.7\\
1064	&	0.61	&	23.0\\
1068	&	0.70	&	57.0\\
1078	&	0.61	&	7.62\\
1087	&	0.59	&	14.6\\
1089	&	0.62	&	26.9\\
1093	&	0.59	&	15.8\\
1095	&	0.61	&	16.6\\
1096	&	0.62	&	8.66\\
1106	&	0.67	&	22.6\\
1112	&	0.69	&	92.1\\
1203	&	0.67	&	36.3\\
1212	&	0.73	&	32.0\\
1218	&	0.63	&	18.5\\
1246	&	0.64	&	20.3\\
1249	&	0.73	&	10.0\\
1251	&	0.70	&	37.5\\
1252	&	0.59	&	17.7\\
1255	&	0.62	&	6.16\\
1258	&	0.61	&	30.8\\
1260	&	0.58	&	13.4\\
1269	&	0.71	&	17.2\\
1278	&	0.73	&	12.0\\
1289	&	0.71	&	8.50\\
1307	&	0.76	&	23.4\\
1341	&	0.62	&	11.7\\
1449	&	0.61	&	14.1\\
1452	&	0.62	&	26.4\\
1462	&	0.63	&	33.8\\
1473	&	0.73	&	22.7\\
1477	&	0.67	&	14.0\\
Sun (1976-04) & 0.65 & 7.90\\
Sun (1976-82) & 0.65 & 9.02\\
Sun (1982-88) & 0.65 & 6.07\\
Sun (1988-94) & 0.65 & 7.49\\
Sun (1994-00) & 0.65 & 5.93\\
Sun (1998-04) & 0.65 & 4.64\\
\enddata
\tablenotetext{a}{Root mean square deviation of the seasonal values of
the HK index}
\end{deluxetable}
\clearpage
\figcaption{Color-magnitude diagram of M67 with the program objects
indicated by filled circles.  Stars in our survey that are known binaries 
are denoted by $\otimes$.  Open circles are proper motion members with 
membership probabilities $\geq$ 90\%}
\figcaption{A comparison of the solar spectrum in the Ca II H and K line region
with the spectrum of an M67 solar-type star (solid line) that has been
calibrated according to the procedure described in $\S$3.  The solar
spectrum (dash line), which is from the {\it NSO Solar Atlas}, has
been degraded to the resolution of the stellar spectrum}
\figcaption{The distribution of the HK index for the M67 solar-type
stars (solid).  Also shown is the HK index distribution for
the contemporary solar cycle,
based on measurements by W. C. Livingston from 1976 to 2004 using the
NSO McMath-Pierce Solar Telescope on Kitt Peak (dash).
The solar data have been adjusted in this comparison to take into account the
effect of the differences in spectral resolution between the solar and
stellar data (see $\S$3 for a discussion).
The solar histogram has been normalized for plotting purposes}
\figcaption{The HK index as a function of intrinsic $B - V$ color for
the solar-type stars in M67.  Note the increase in scatter toward
cooler stars}
\figcaption{A sequence of histograms showing the distribution of HK index
in bins of $B - V$ color for the 60 solar-type stars observed in M67.
The solar cycle distribution (dash) is provided for comparative purposes. See
$\S$4 for a discussion}
\figcaption{The HK index distribution of known binaries ({\it solid}) and stars
that are not known to be binary systems ({\it dash}) in our M67 sample of
solar-type stars. See $\S$4 for a discussion}
\figcaption{The HK index versus orbital period for the binaries in our
sample.  The periods were kindly provided in advance of publication by
R. D. Mathieu}
\figcaption{The cumulative probability functions in HK index for the 
solar-type stars in M67 ({\it solid}) and the modern solar cycle ({\it dash}), 
respectively.  See $\S$5 for a discussion}
\figcaption{The seasonal variation of chromospheric activity in solar-type
stars in M67.  The Sun seen as a star is included for comparison.  See $\S$5.2 
for a discussion}
\figcaption{The appearance of the solar cycle variation in annual mean HK
index.  The solar HK index values are based on measurements
provided by W. C. Livingston in advance of publication, and have been adjusted
for the purposes of comparison with the M67 stellar data as discussed
in $\S$3. The dotted vertical
grid lines are drawn at 6-year intervals to facilitate comparison of
segments of the solar cycle of length similar to that of the record of seasonal
variations of the M67 stars (Fig. 9)}
\figcaption{The root mean square deviation of the seasonal values of the HK 
index vs. mean HK index for the M67 solar-type stars in Table 2. See $\S$5.7 
for a discussion}
\figcaption{The distribution of the seasonal values of the HK index (solid)
along with the distribution of values seen in the modern solar cycle}
\figcaption{The apparent age distribution as inferred from the current 
calibration of chromospheric activity and age for the M67 solar-type stars
({\it solid}).  The corresponding distribution for the solar cycle is 
shown for comparison ({\it dash}). See $\S$5.3 for a discussion}
\figcaption{An example of an especially active solar-type star in M67.  The
emission reversals in the Ca II H and K cores are clearly evident in this
object}
%
\begin{figure}
\figurenum{1}
\epsscale{}
\includegraphics*[scale=0.90,angle=180]{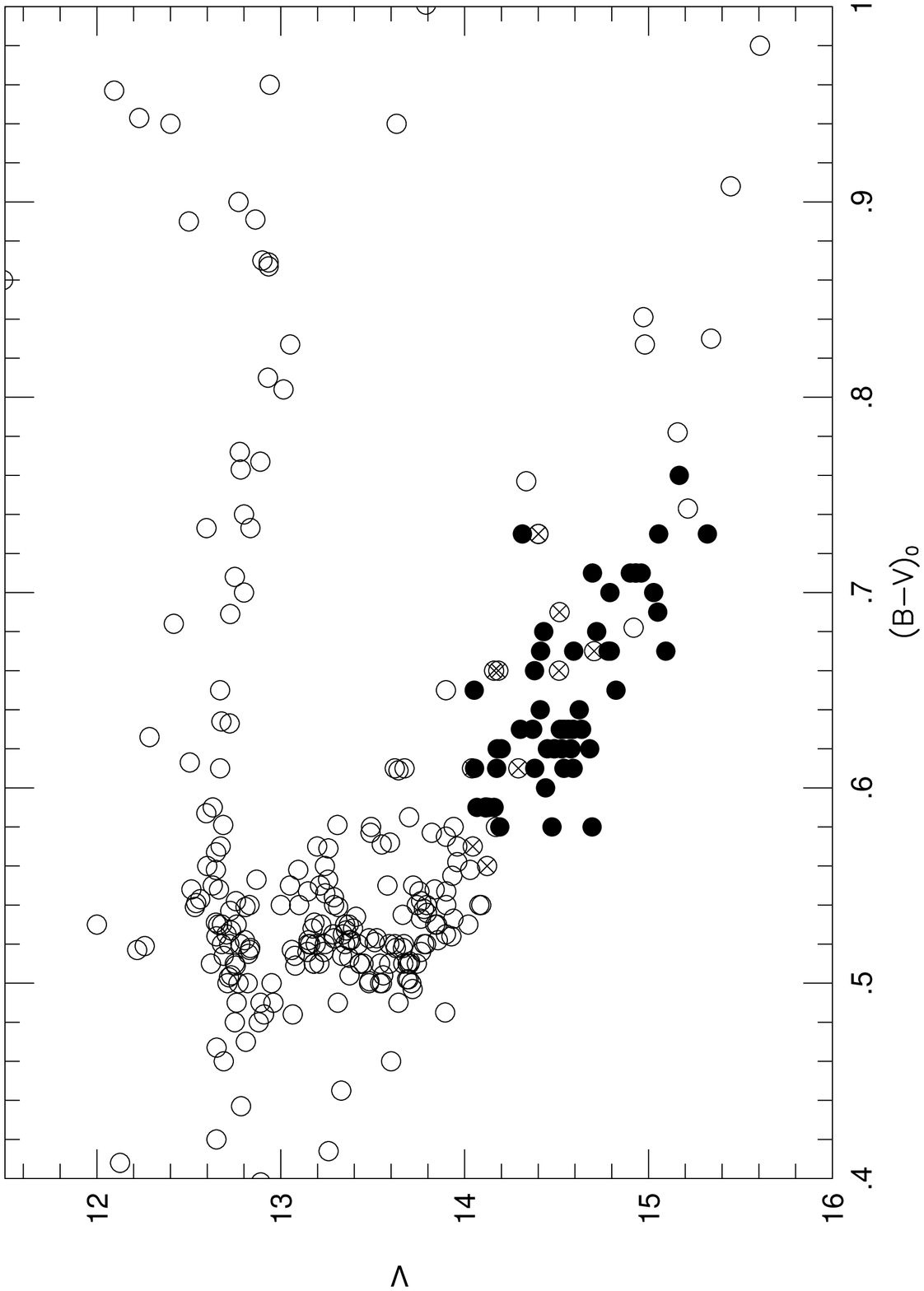}
\end{figure}
\begin{figure}
\figurenum{2}
\includegraphics*[scale=0.90,angle=180]{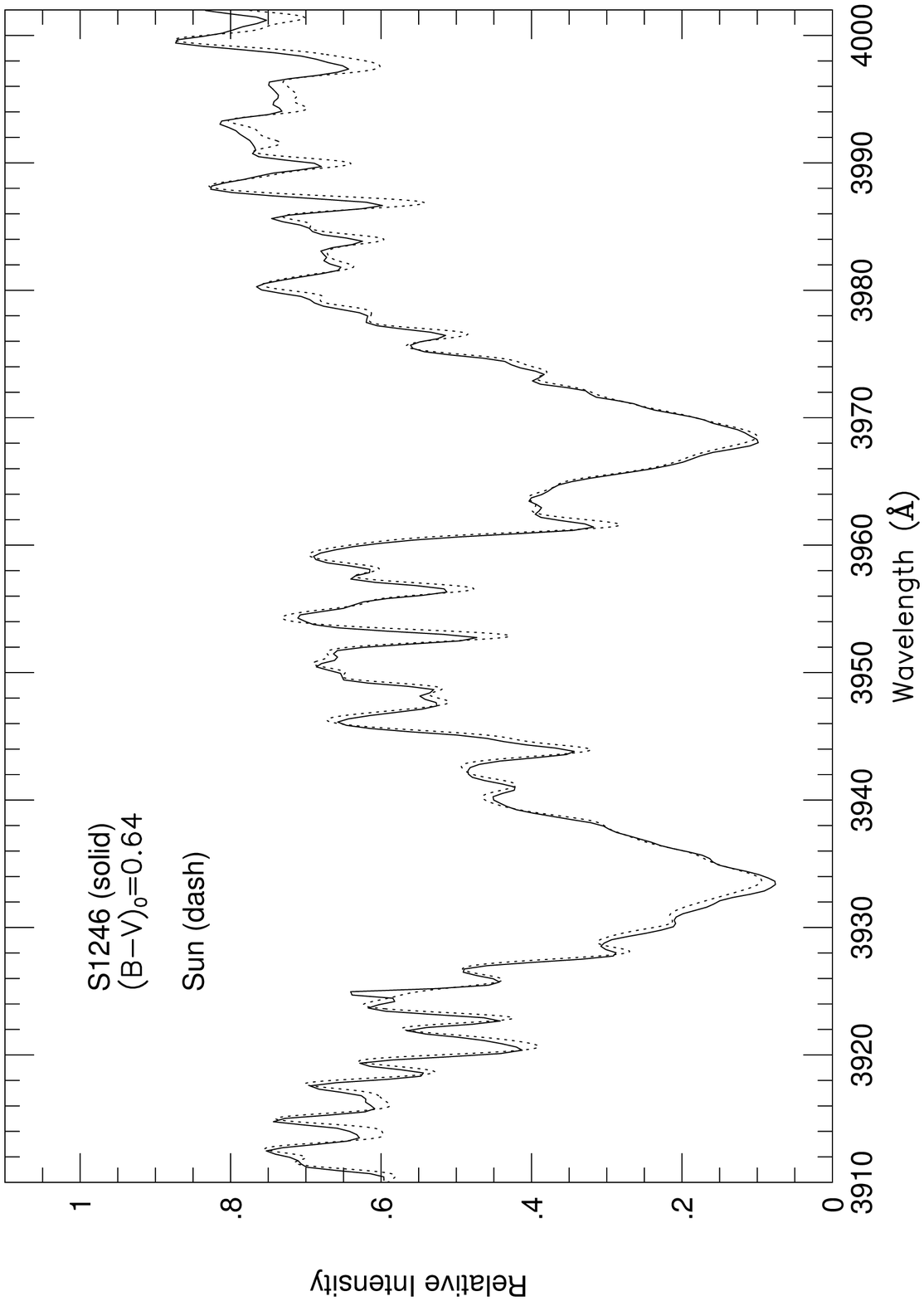}
\end{figure}
\begin{figure}
\figurenum{3}
\includegraphics*[scale=0.90,angle=180]{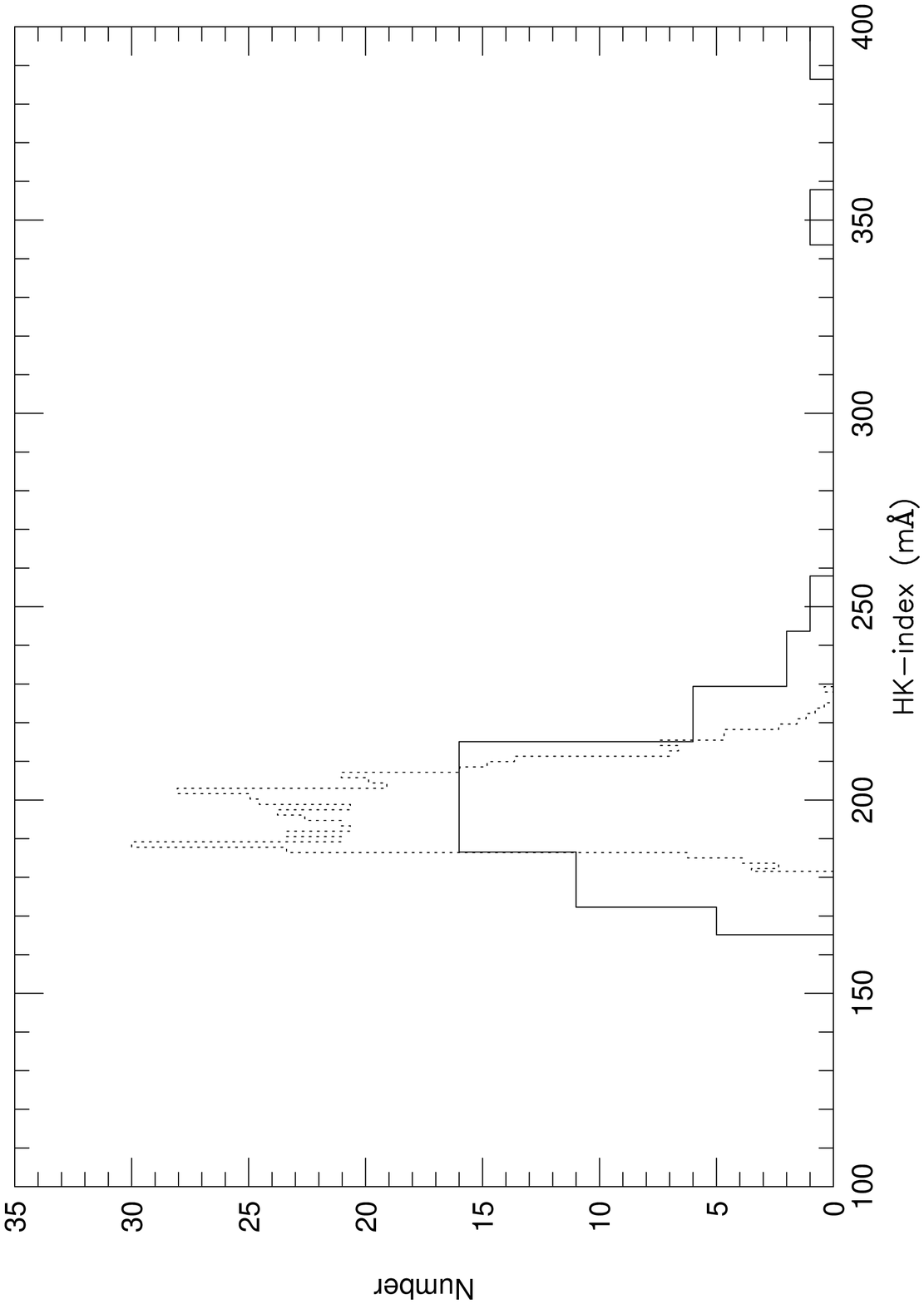}
\end{figure}
\begin{figure}
\figurenum{4}
\includegraphics*[scale=0.90,angle=180]{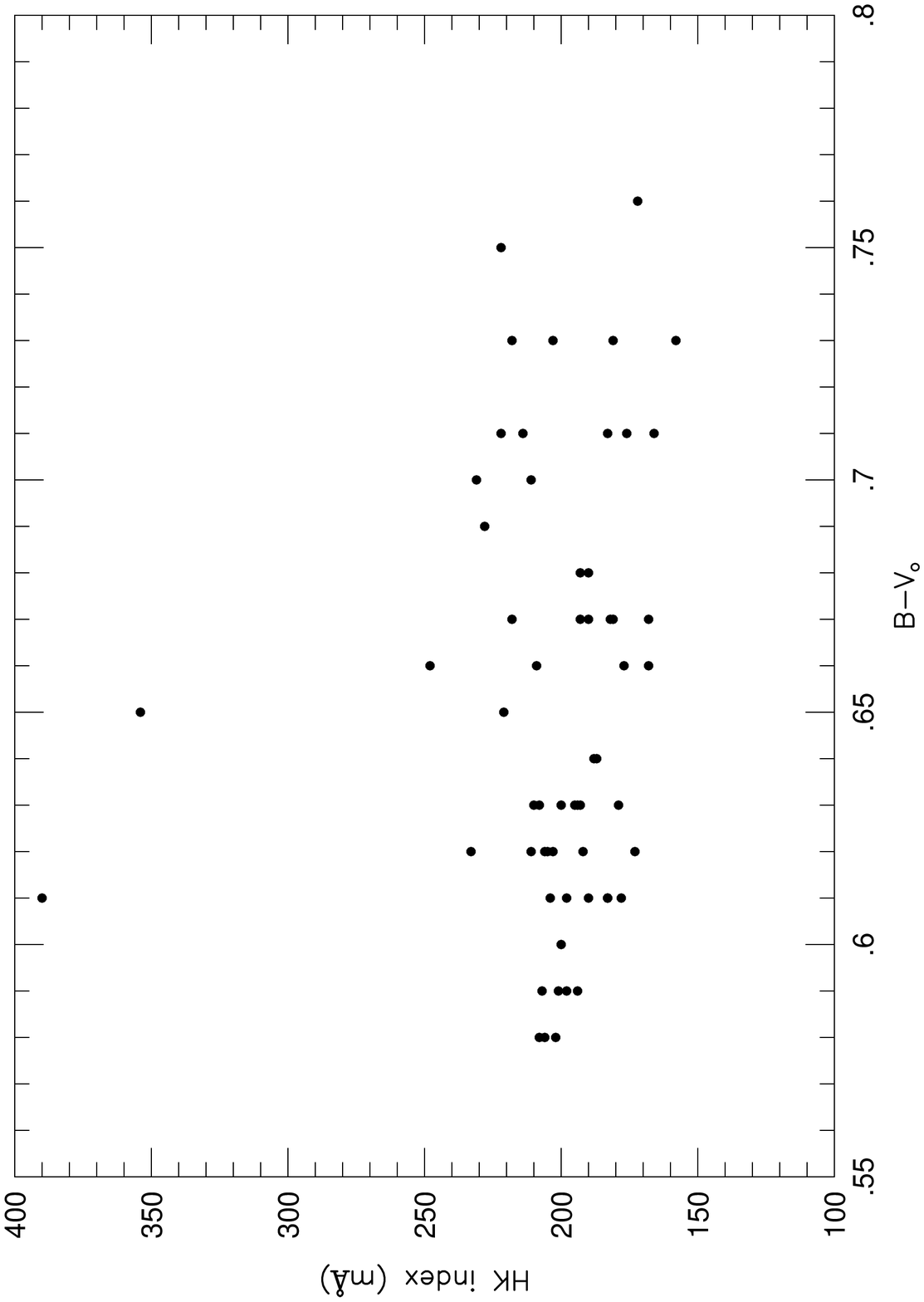}
\end{figure}
\begin{figure}
\figurenum{5}
\includegraphics*[scale=0.90,angle=180]{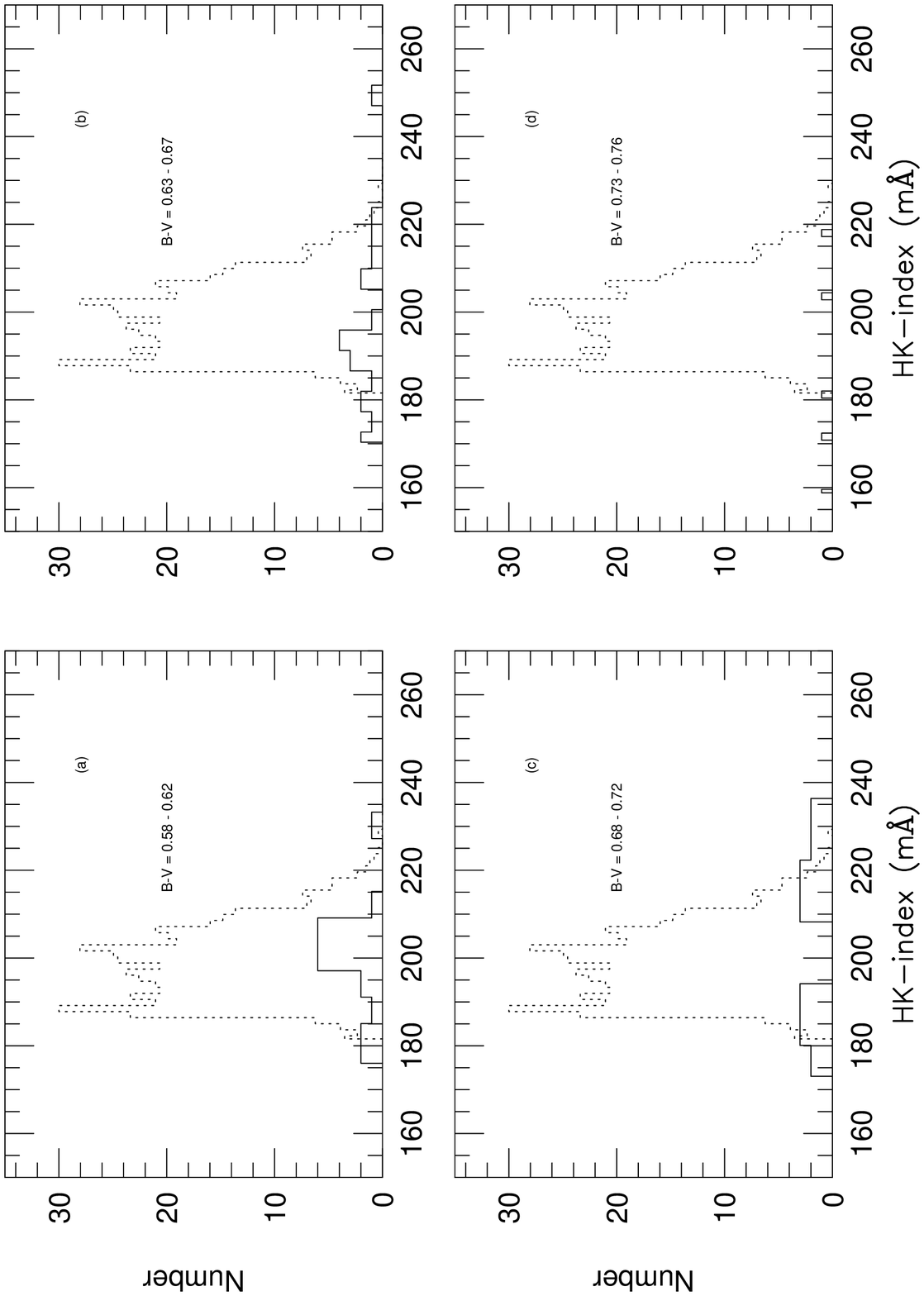}
\end{figure}
\begin{figure}
\figurenum{6}
\includegraphics*[scale=0.90,angle=180]{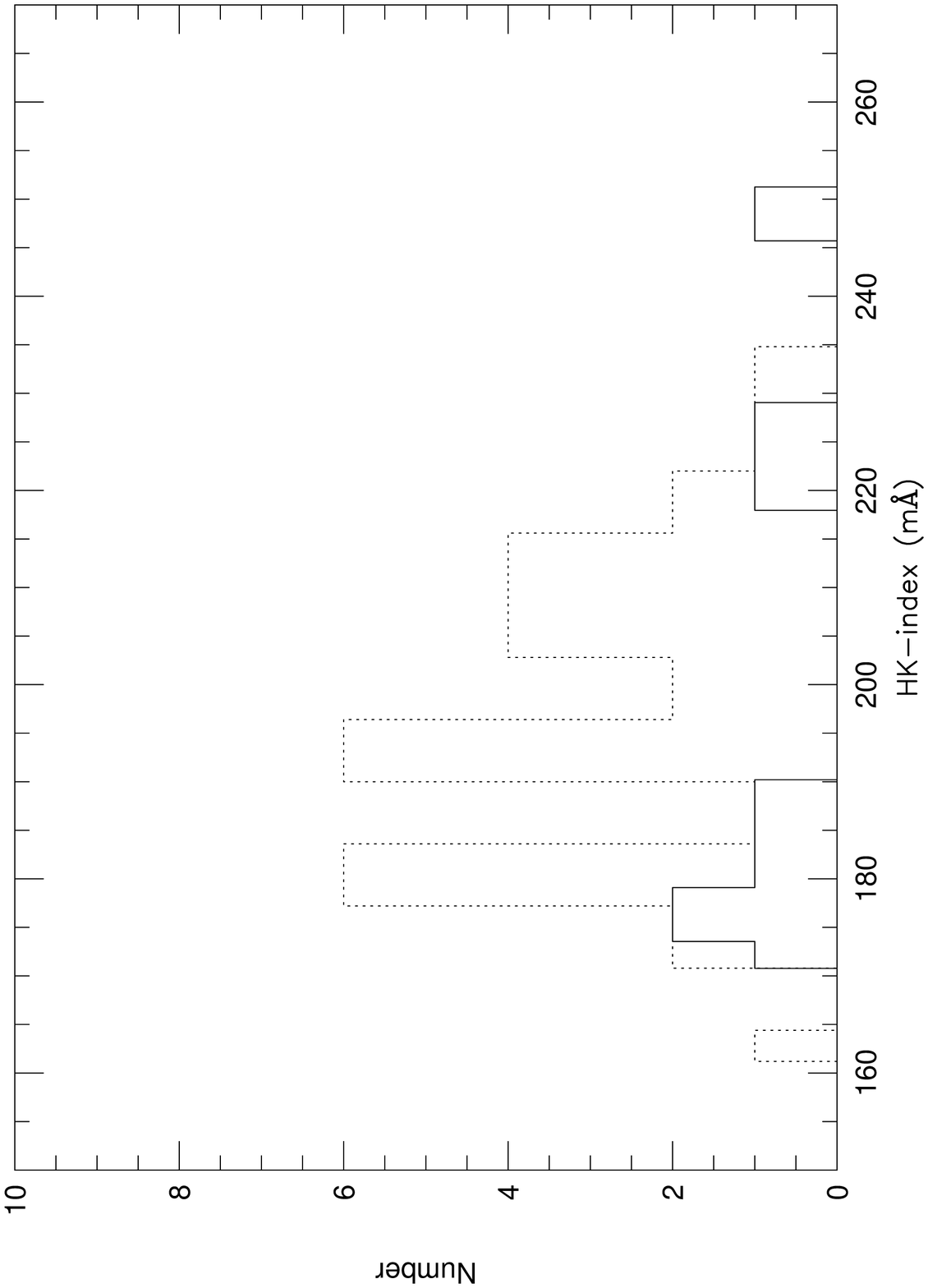}
\end{figure}
\begin{figure}
\figurenum{7}
\includegraphics*[scale=0.90,angle=180]{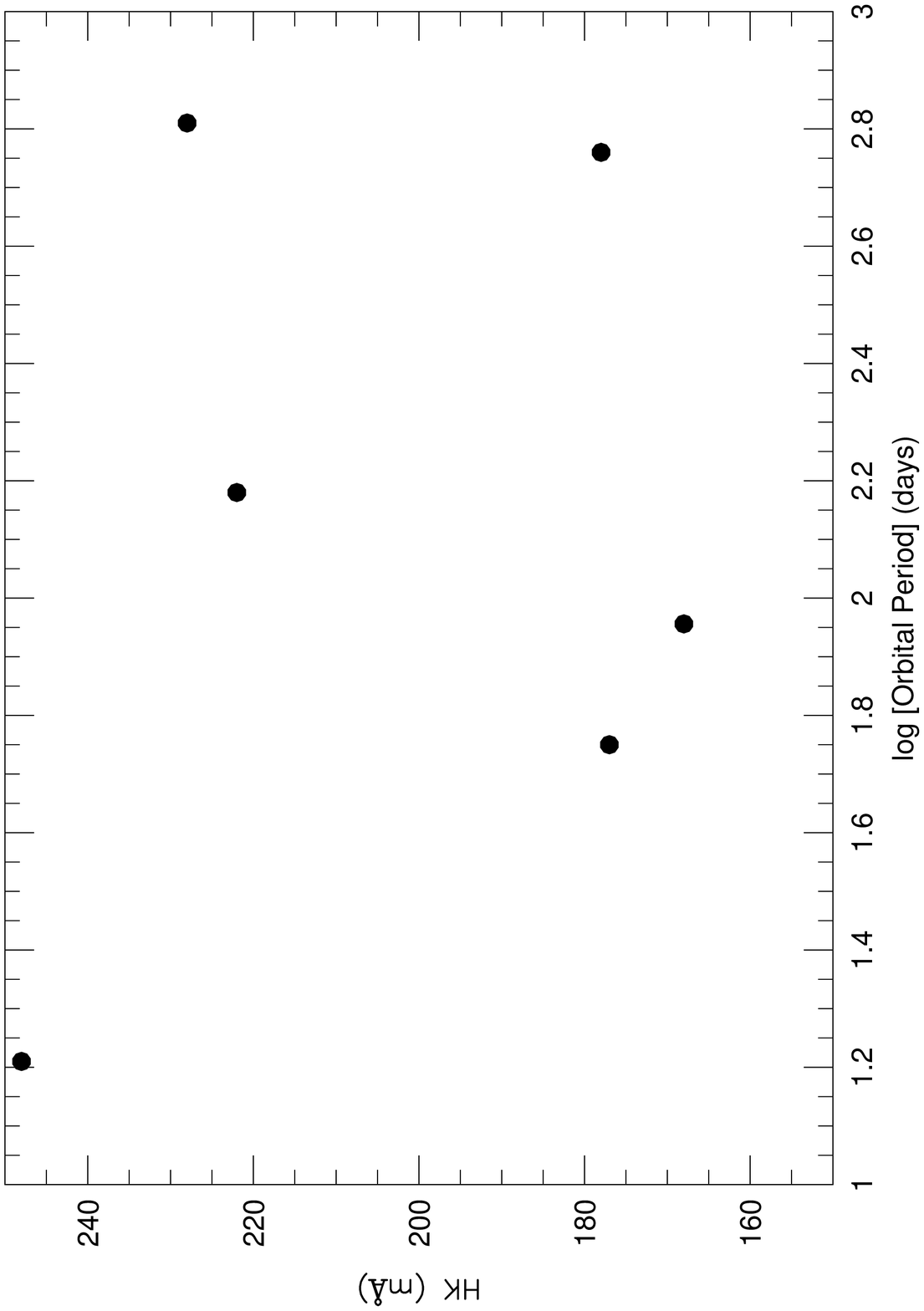}
\end{figure}

\begin{figure}
\figurenum{8}
\includegraphics*[scale=0.90,angle=180]{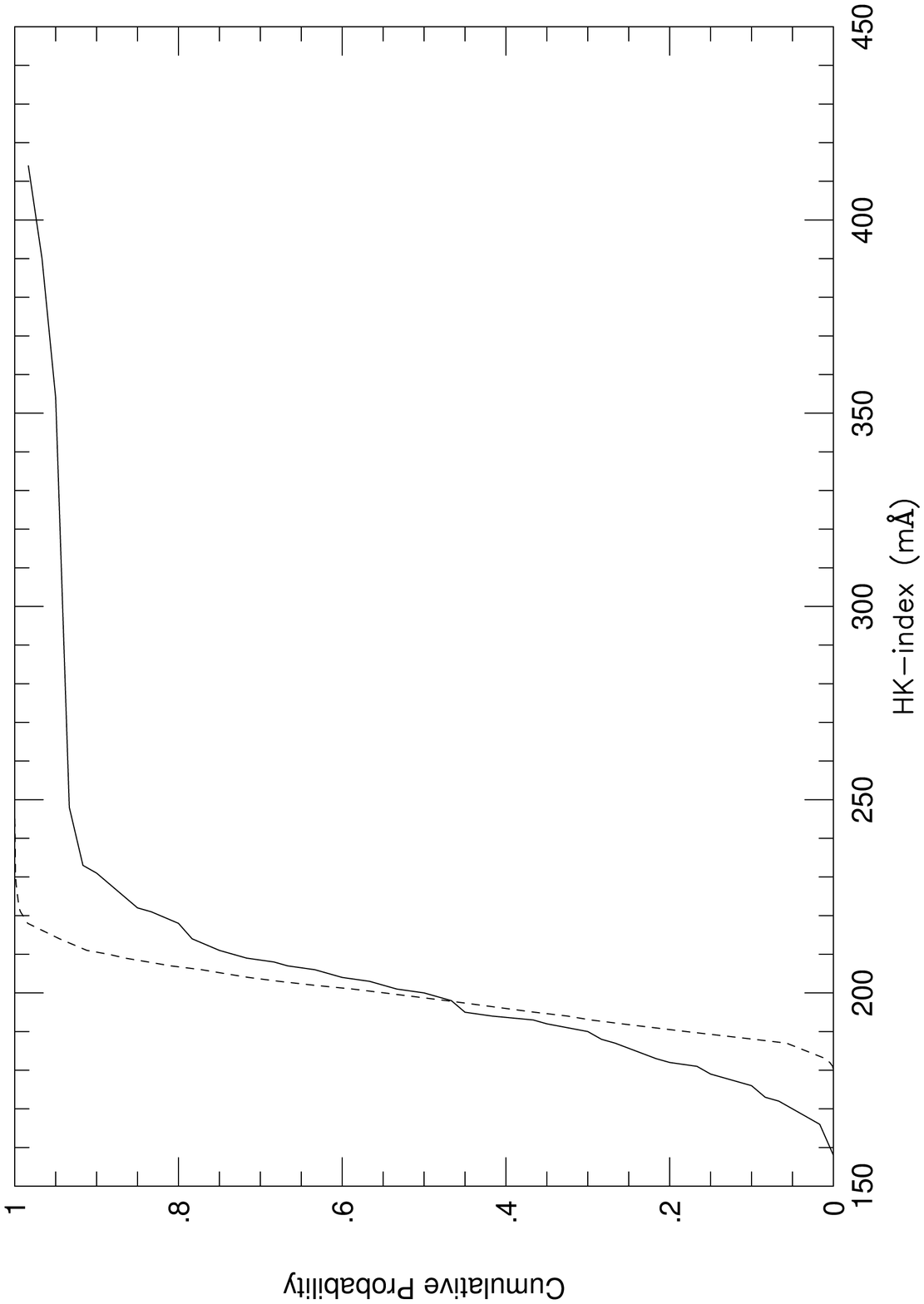}
\end{figure}
%
\begin{figure}
\figurenum{9}
\includegraphics*[scale=0.90,angle=180]{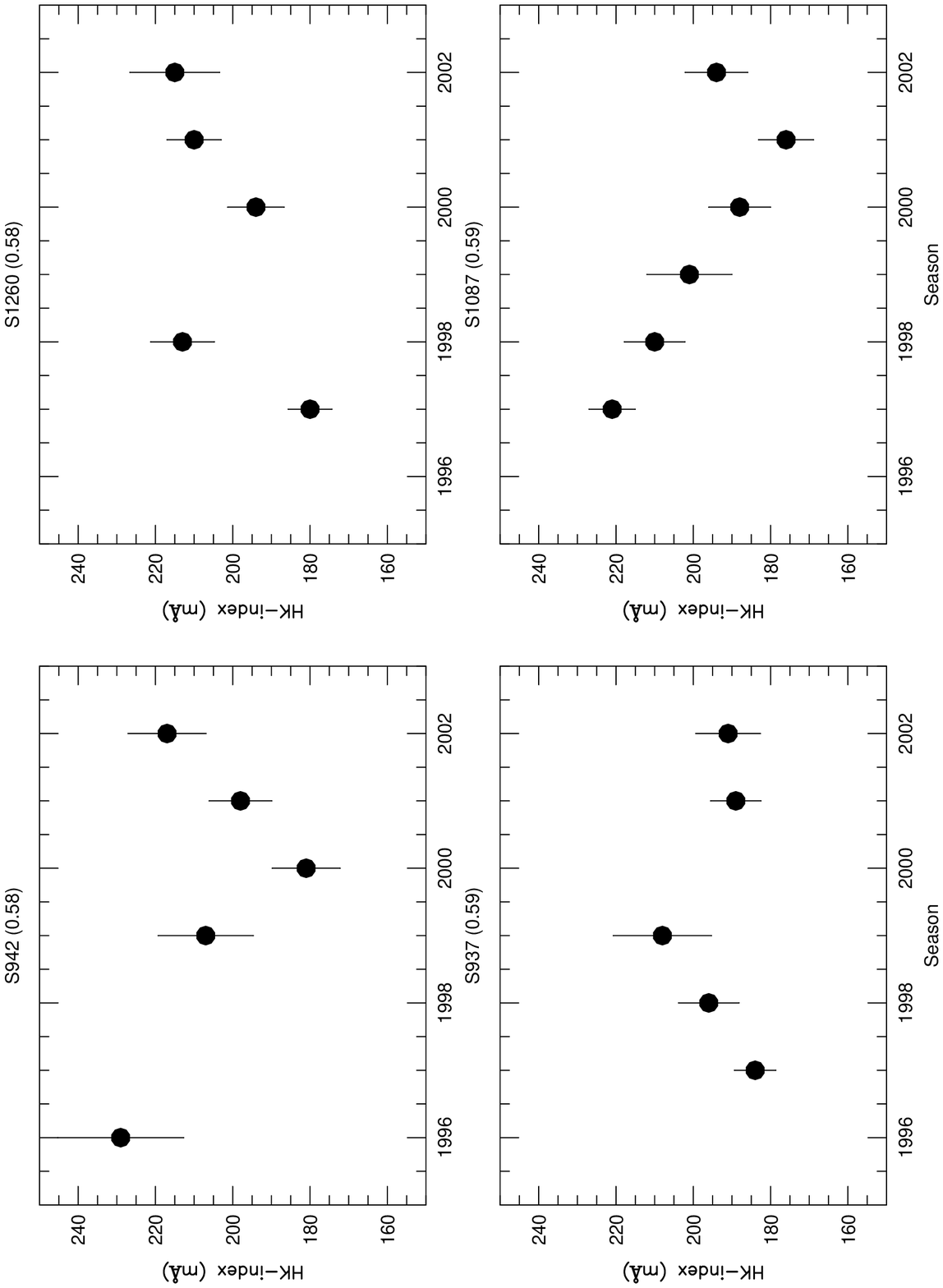}
\end{figure}
\begin{figure}
\includegraphics*[scale=0.90,angle=180]{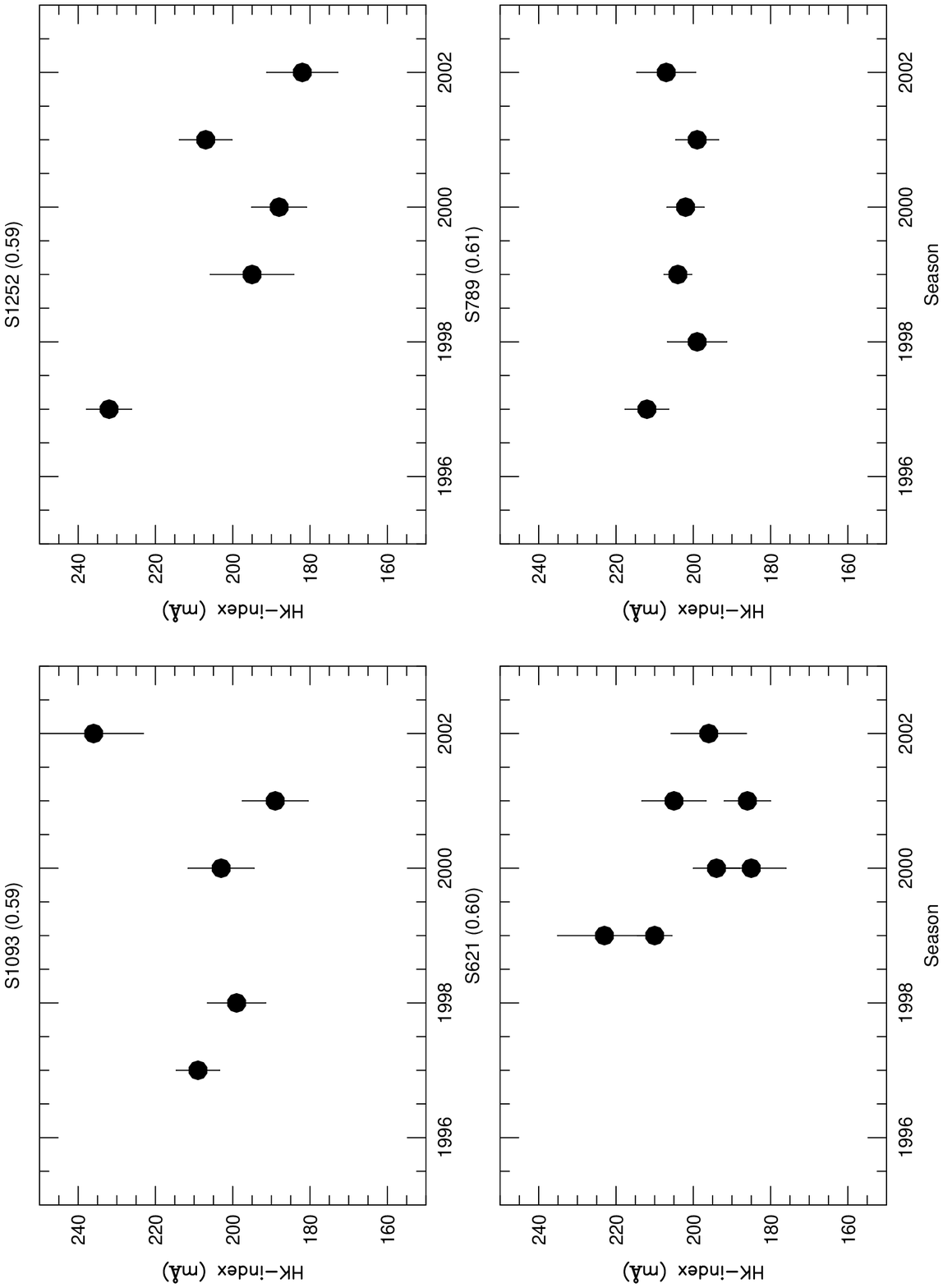}
\end{figure}
\begin{figure}
\includegraphics*[scale=0.90,angle=180]{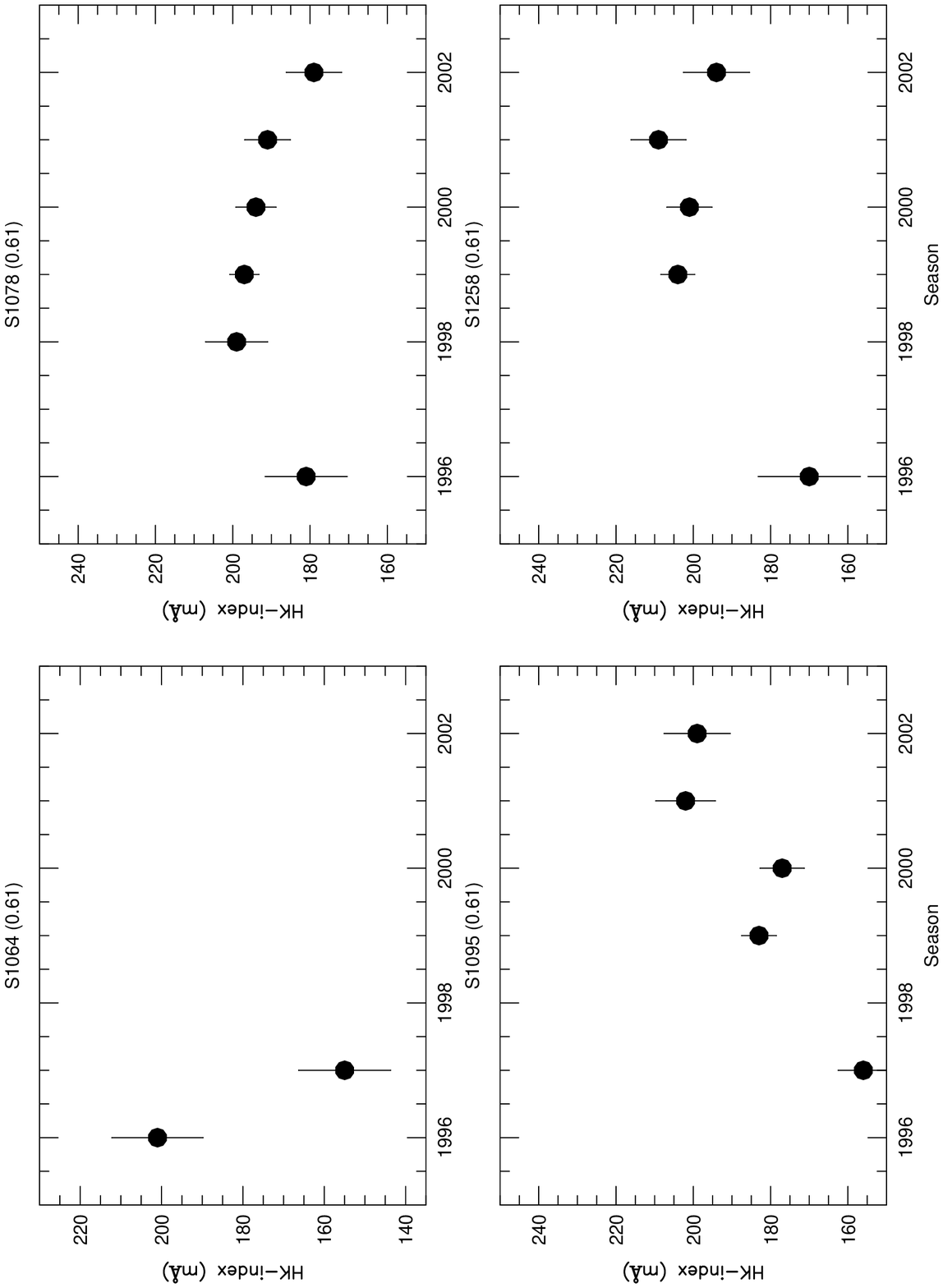}
\end{figure}
\begin{figure}
\includegraphics*[scale=0.90,angle=180]{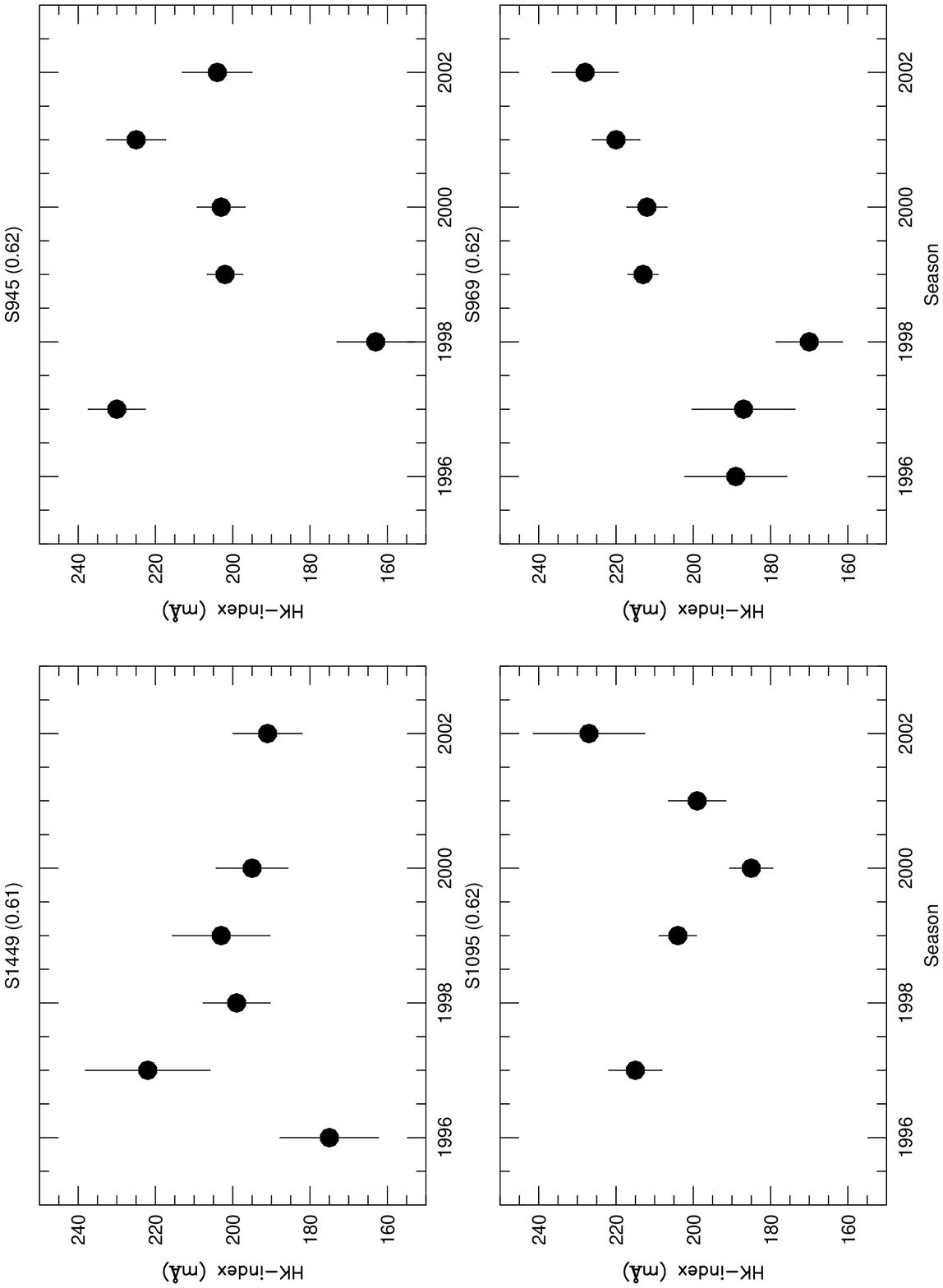}
\end{figure}
\begin{figure}
\includegraphics*[scale=0.90,angle=180]{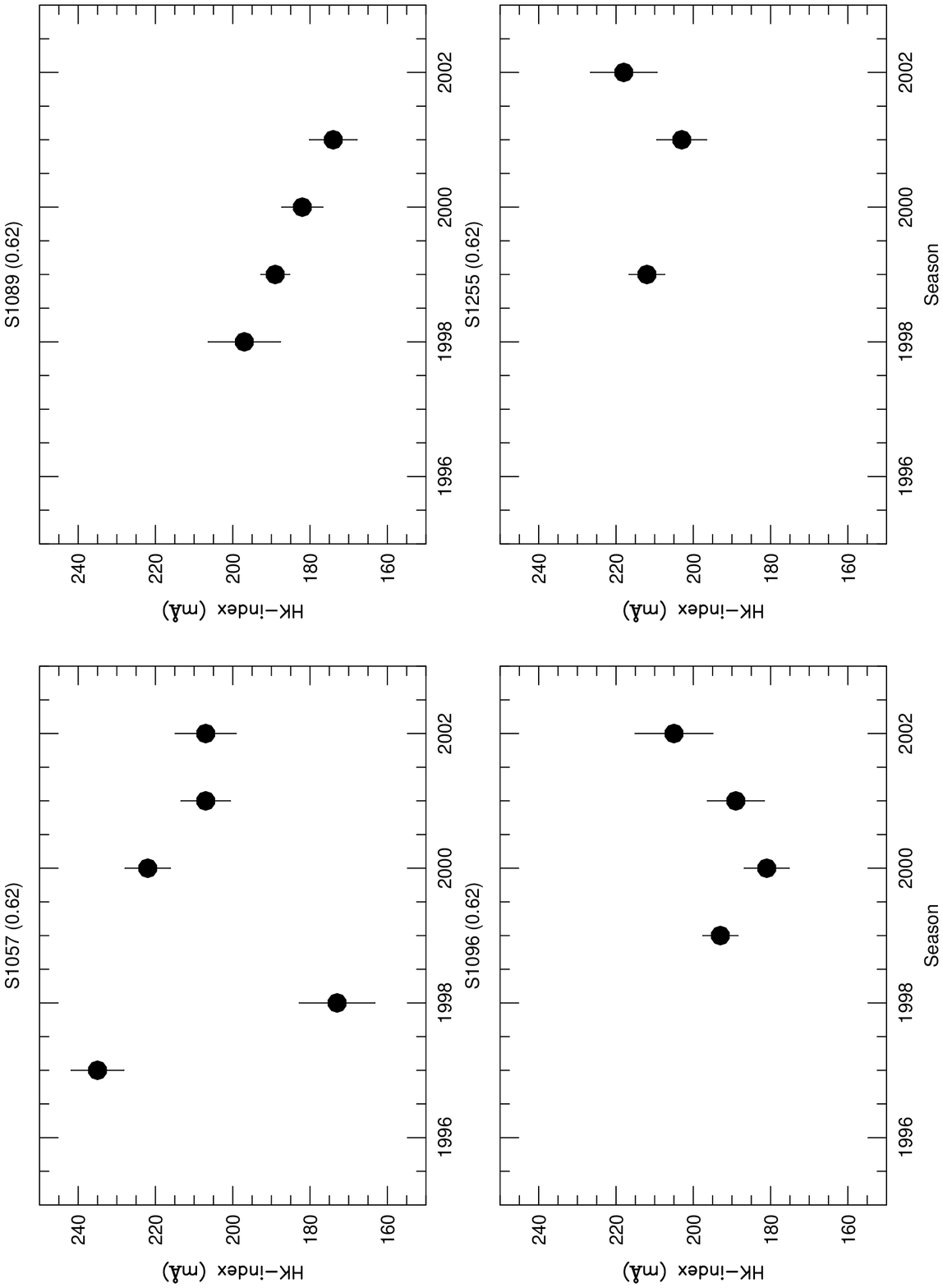}
\end{figure}
\begin{figure}
\includegraphics*[scale=0.90,angle=180]{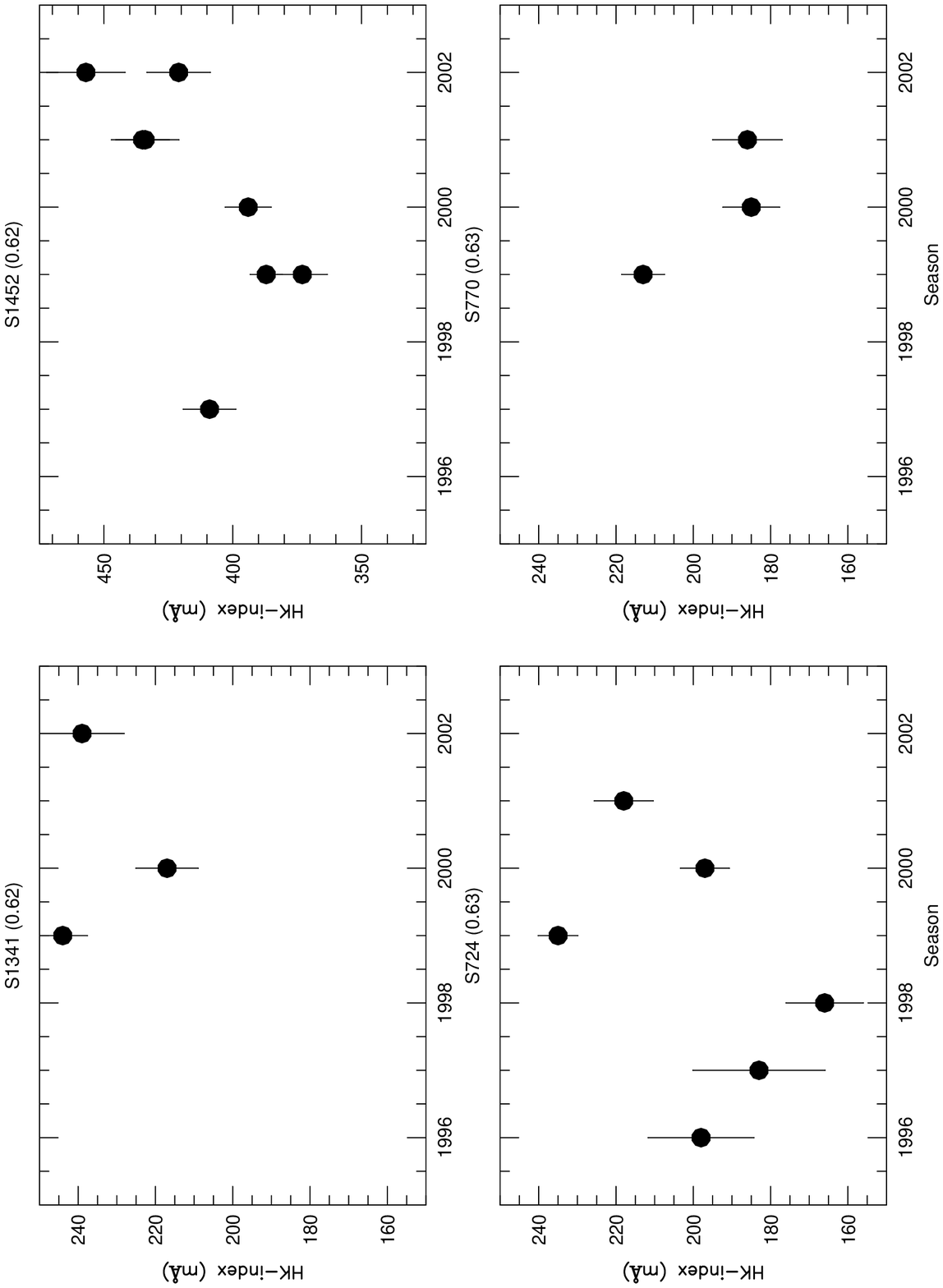}
\end{figure}
\begin{figure}
\includegraphics*[scale=0.90,angle=180]{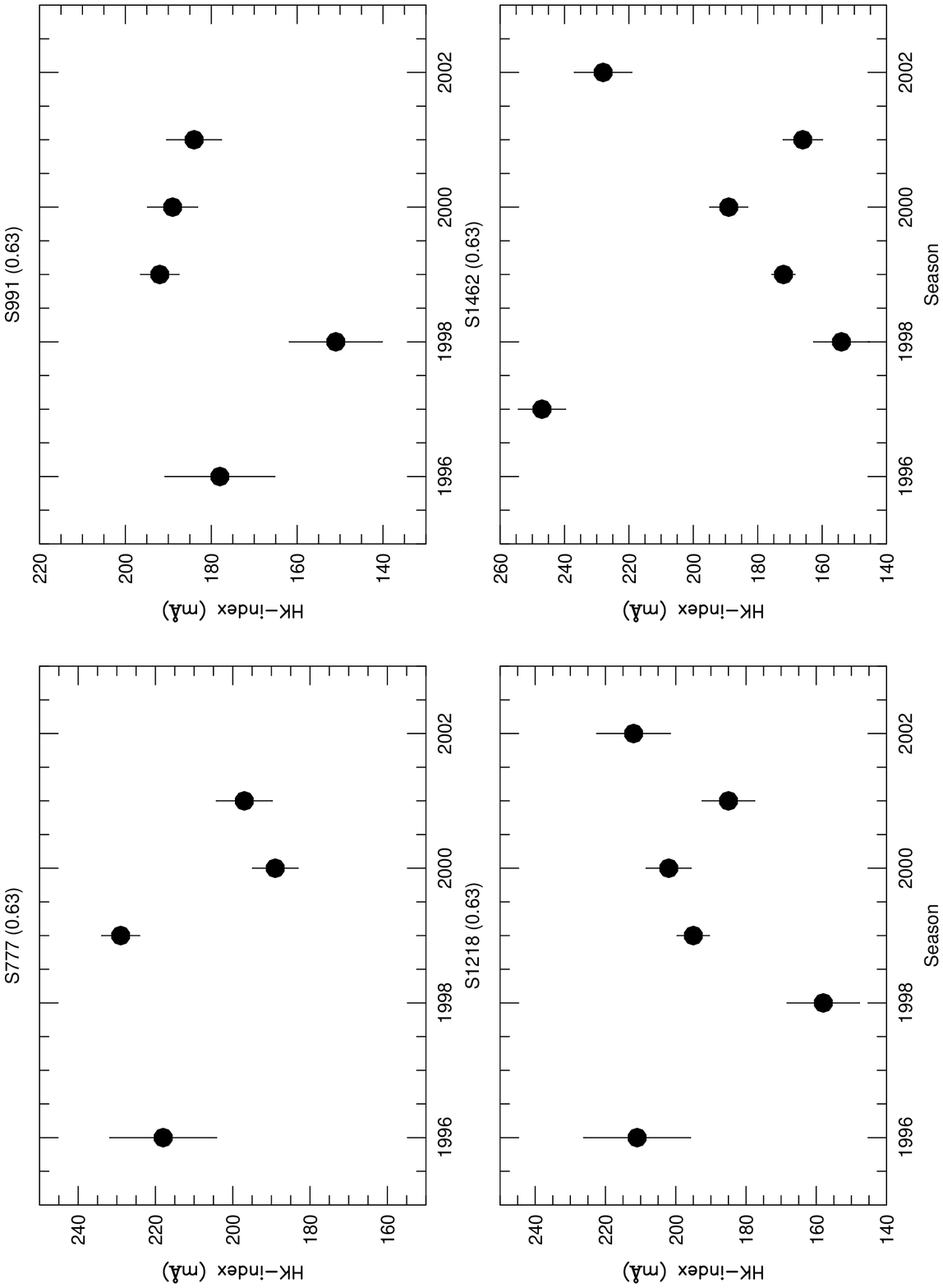}
\end{figure}
\begin{figure}
\includegraphics*[scale=0.90,angle=180]{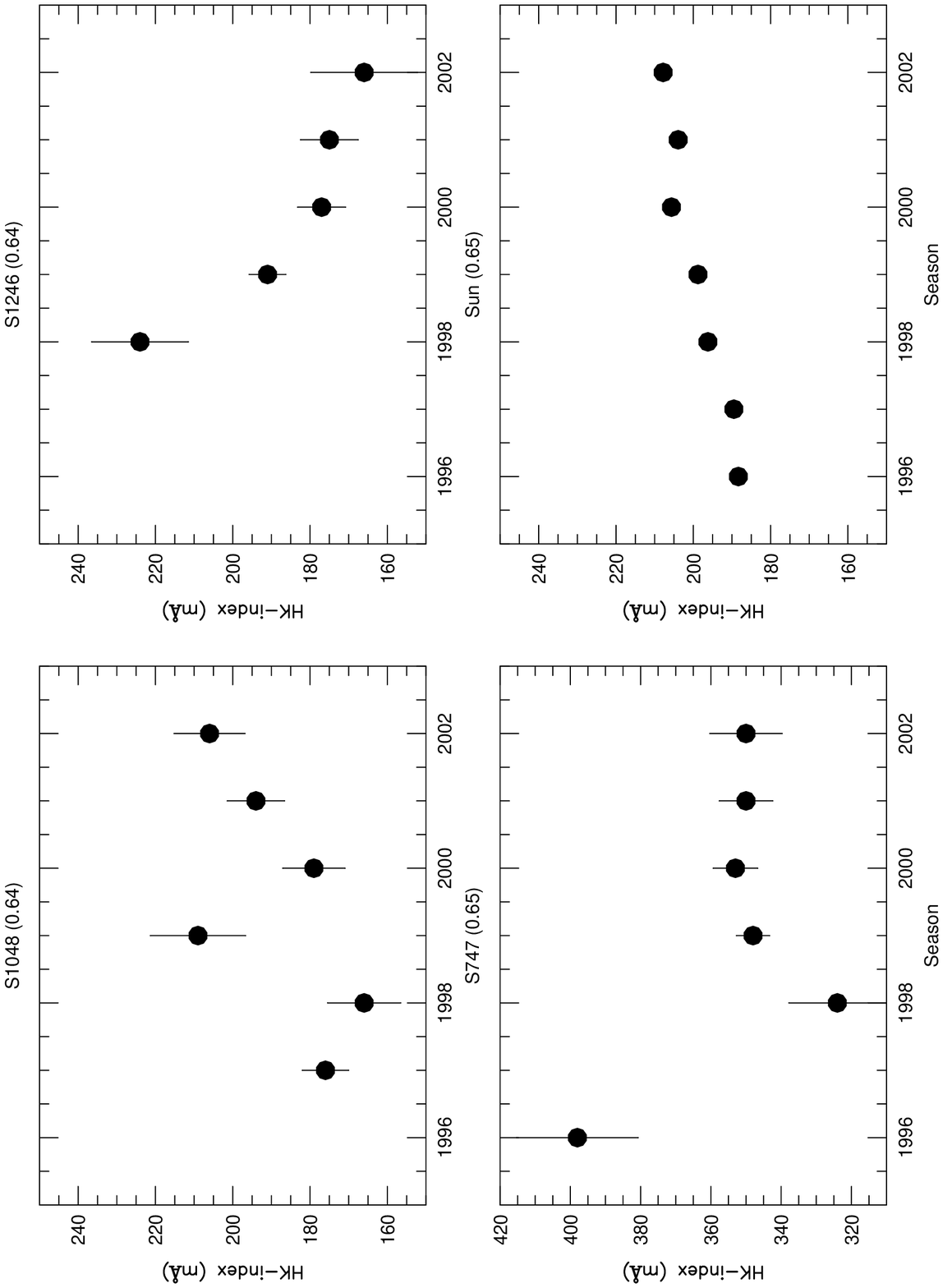}
\end{figure}
\begin{figure}
\includegraphics*[scale=0.90,angle=180]{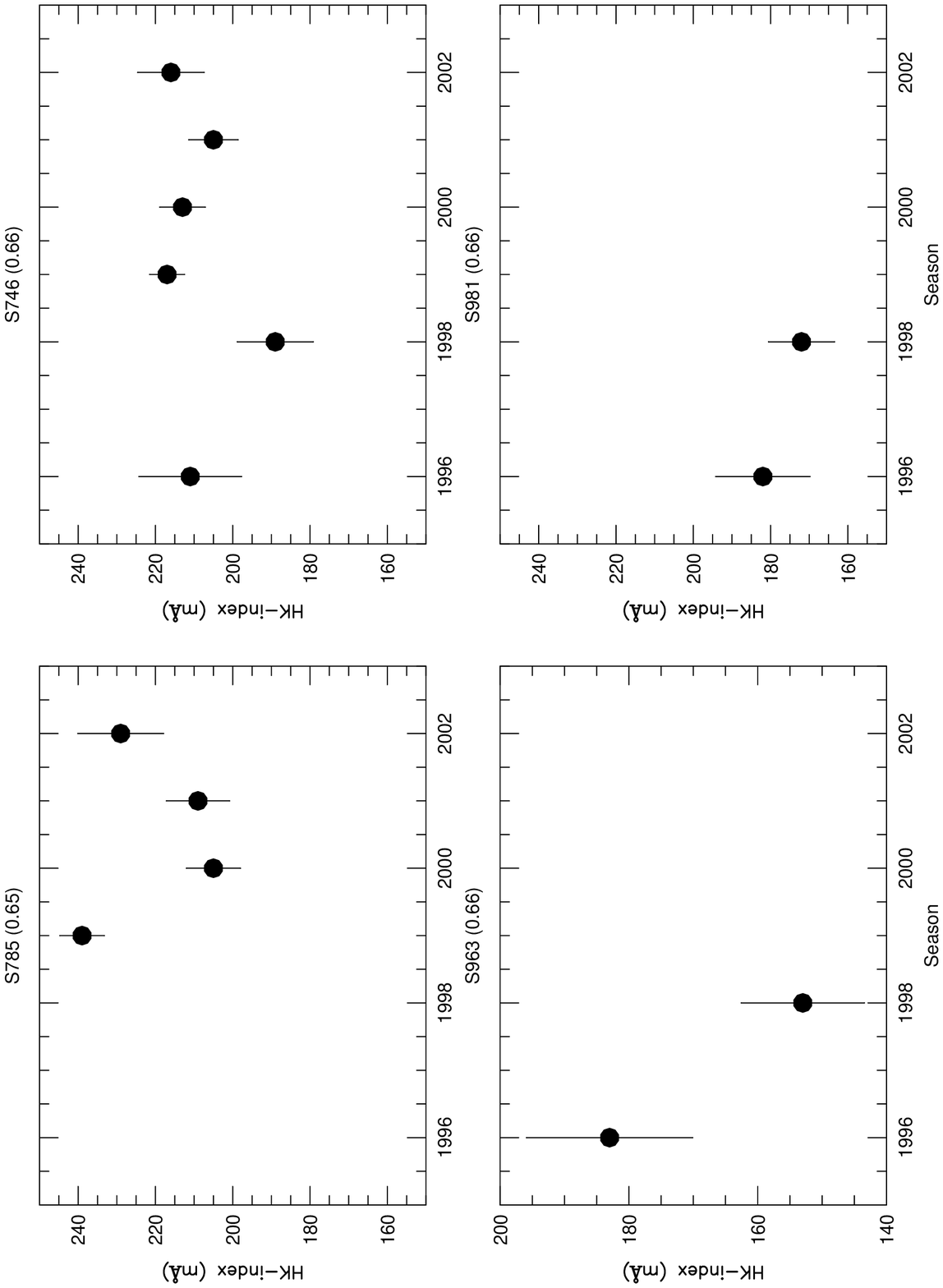}
\end{figure}

\clearpage

\begin{figure}
\includegraphics*[scale=0.90,angle=180]{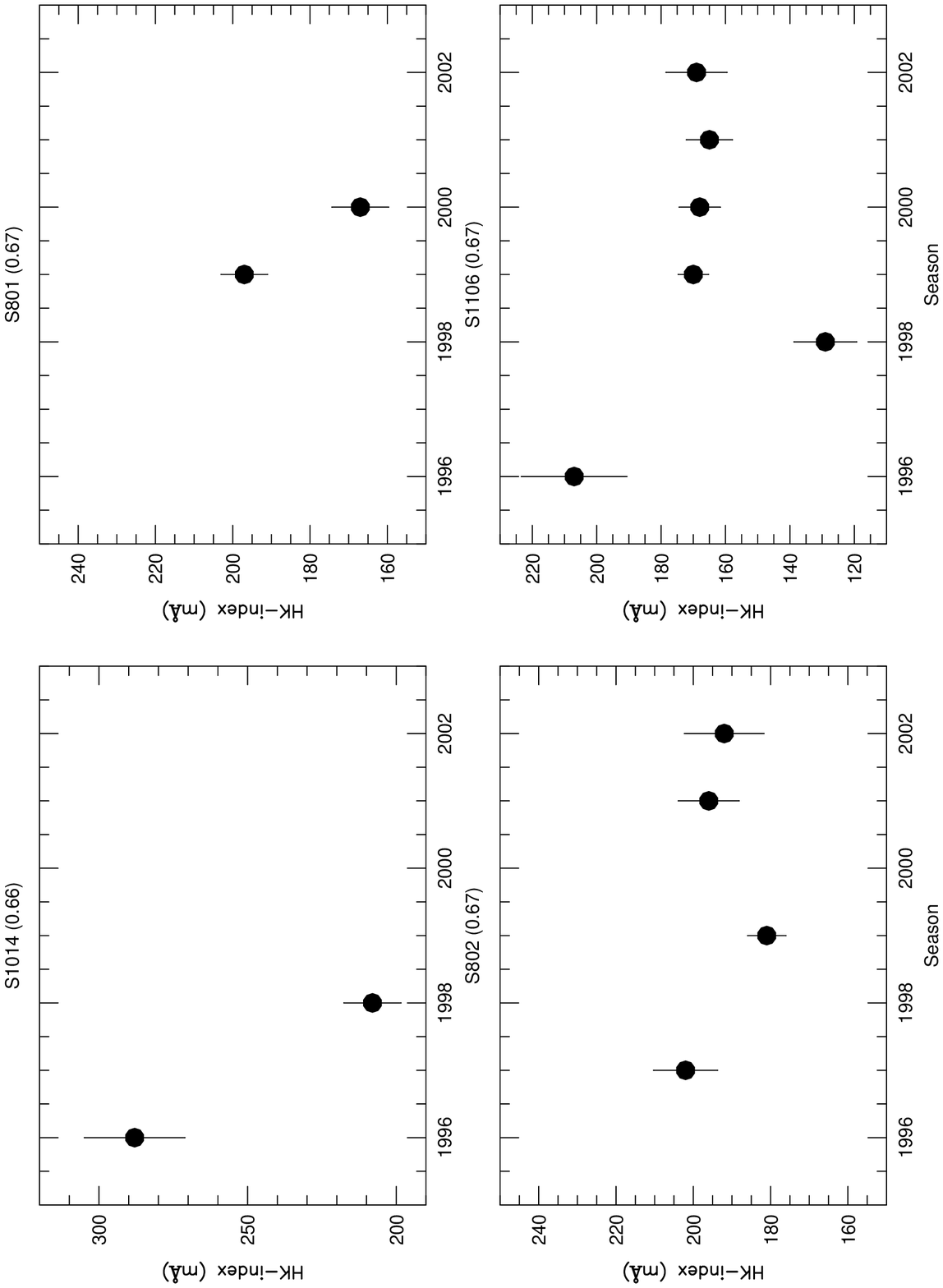}
\end{figure}
\begin{figure}
\includegraphics*[scale=0.90,angle=180]{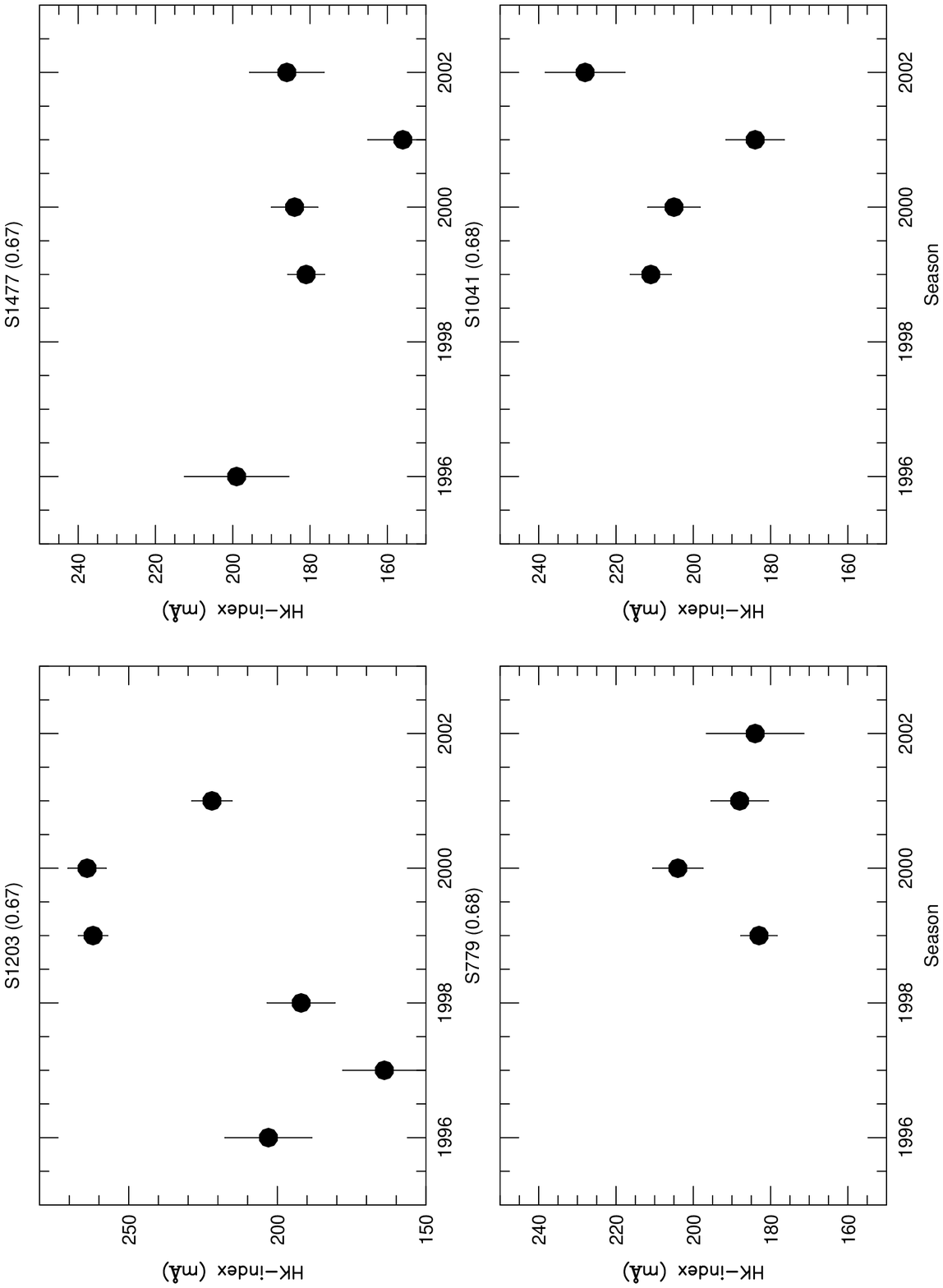}
\end{figure}
\begin{figure}
\includegraphics*[scale=0.90,angle=180]{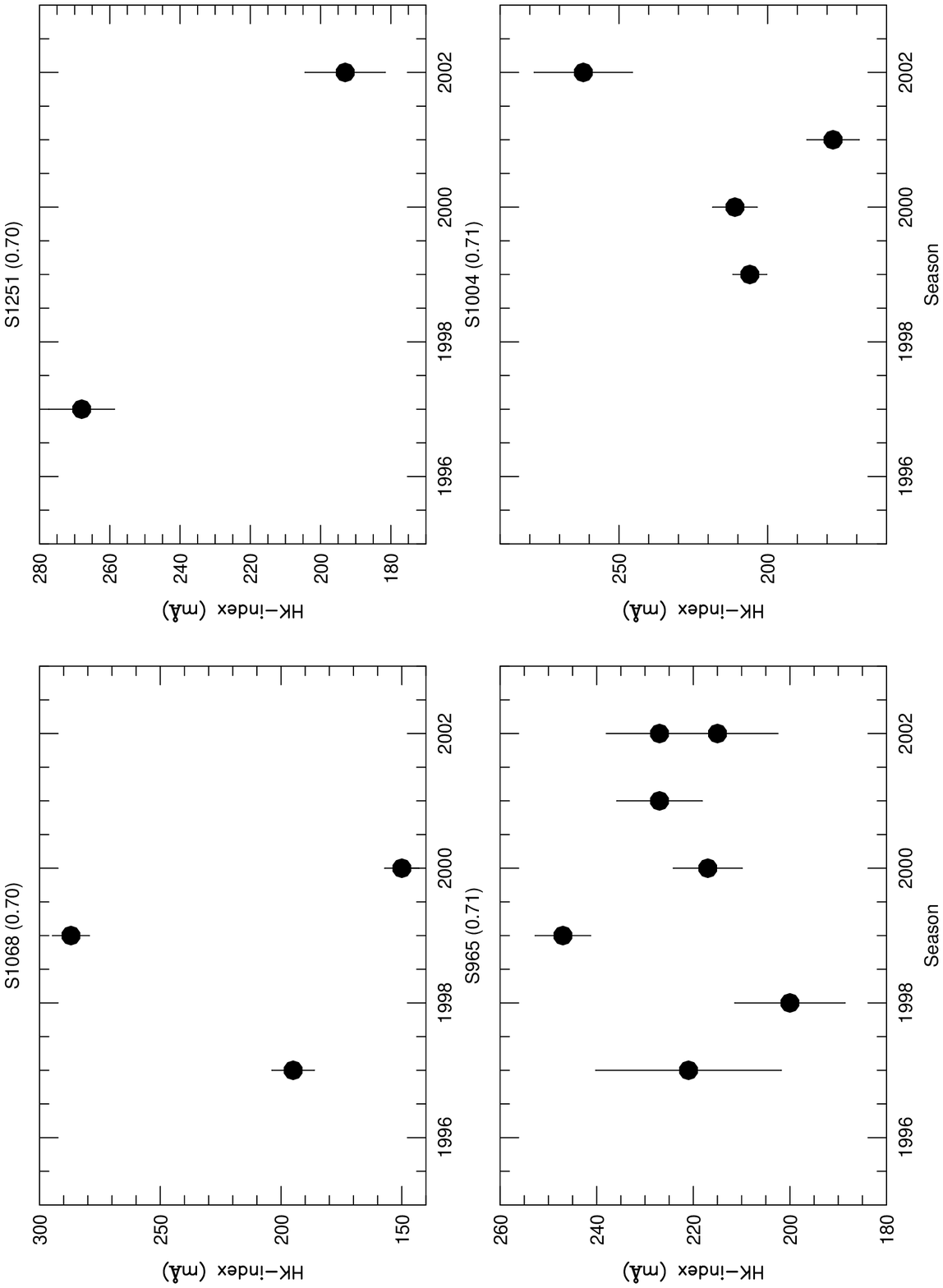}
\end{figure}
\begin{figure}
\includegraphics*[scale=0.90,angle=180]{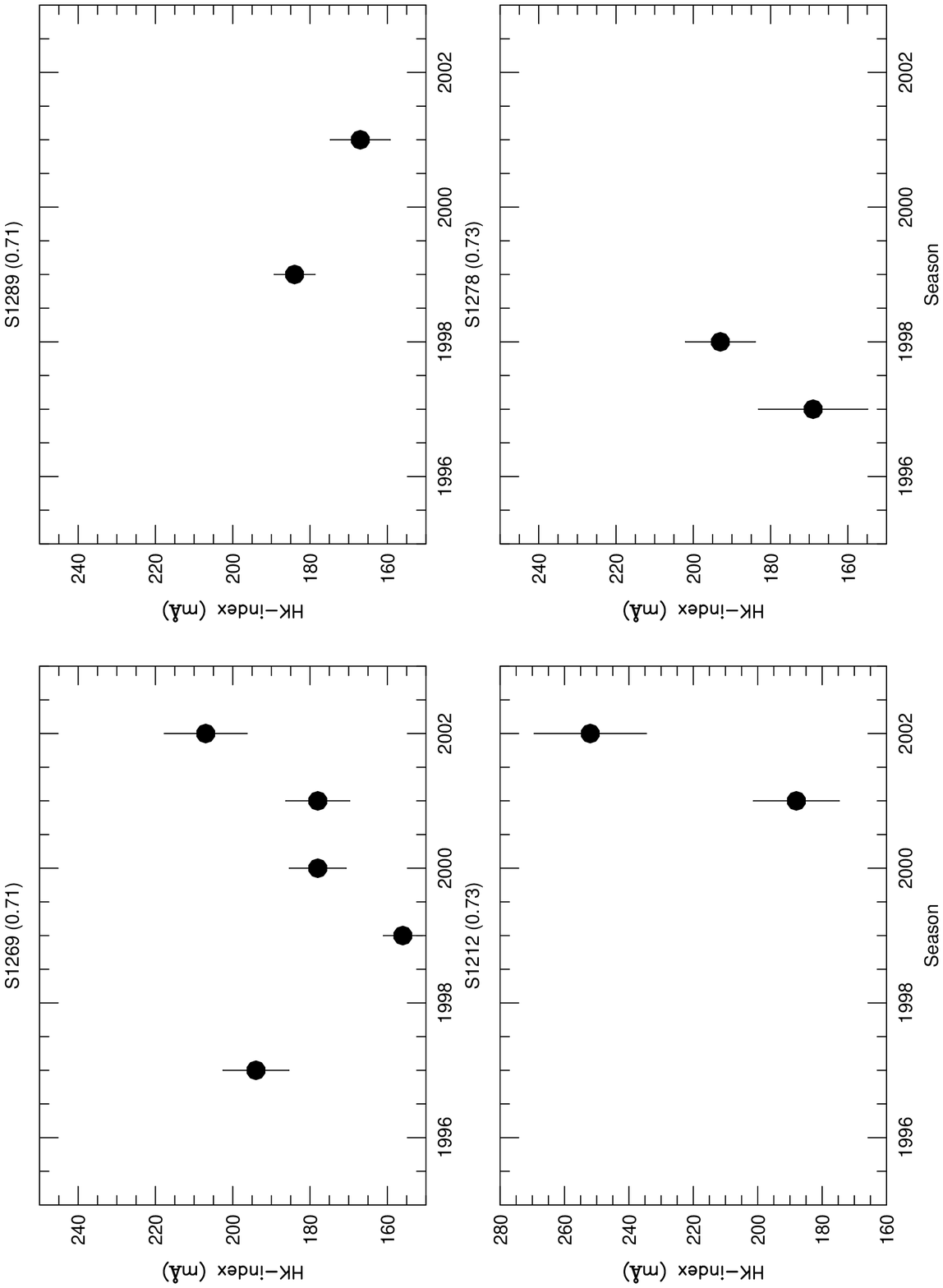}
\end{figure}
\begin{figure}
\includegraphics*[scale=0.90,angle=180]{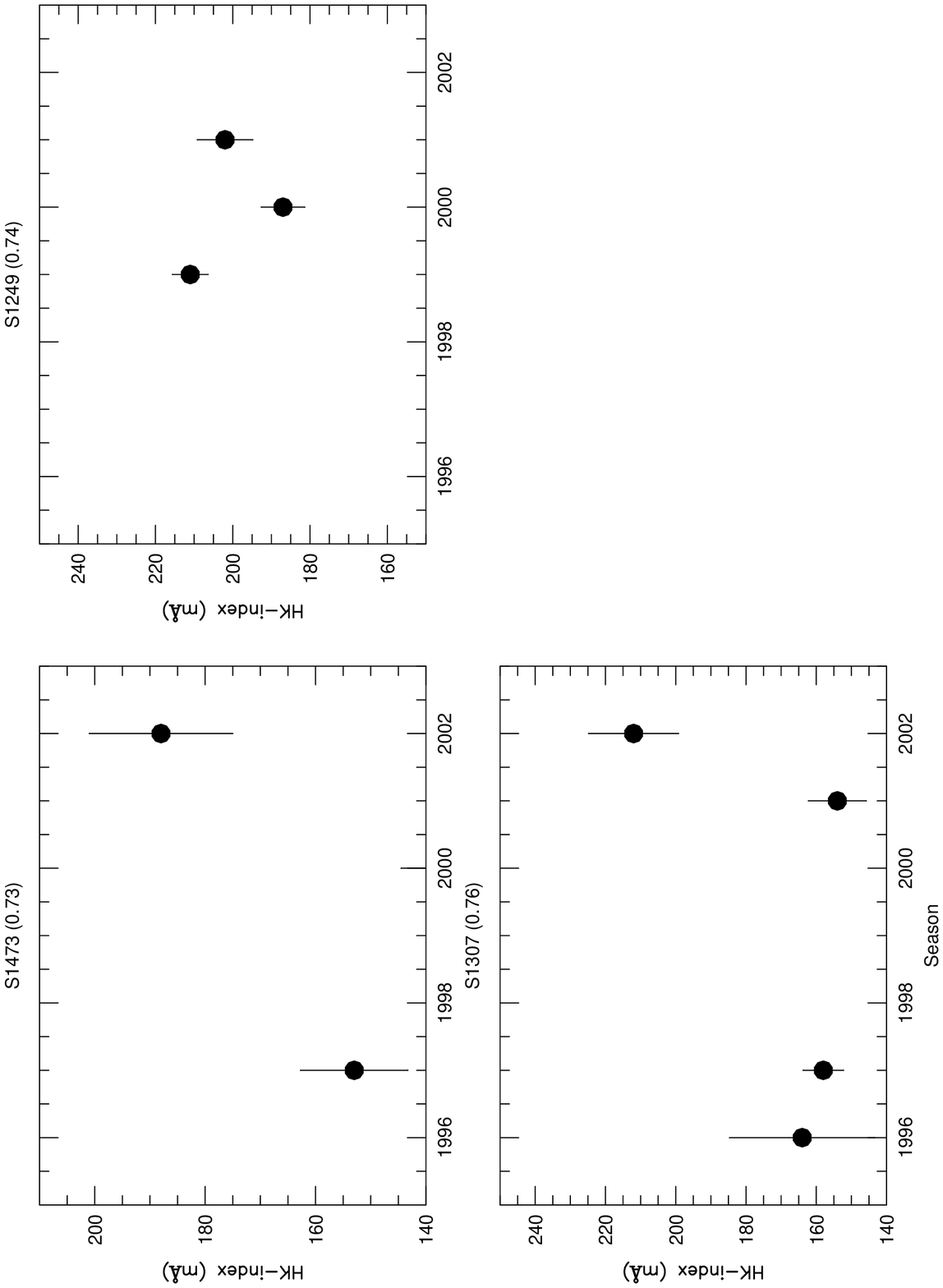}
\end{figure}

%
%
\begin{figure}
\figurenum{10}
\includegraphics*[scale=0.90,angle=180]{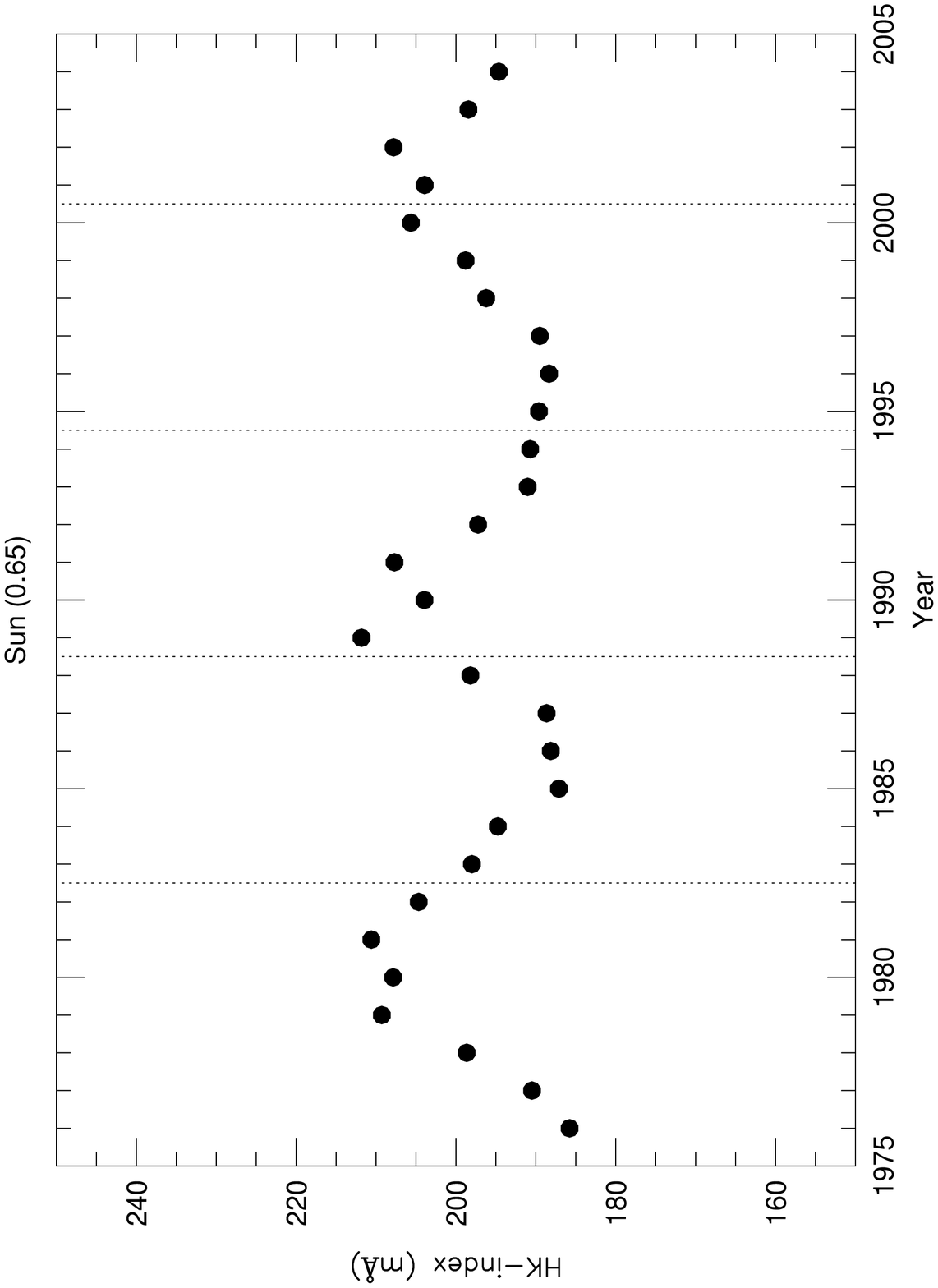}
\end{figure}

\begin{figure}
\figurenum{11}
\includegraphics*[scale=0.90,angle=180]{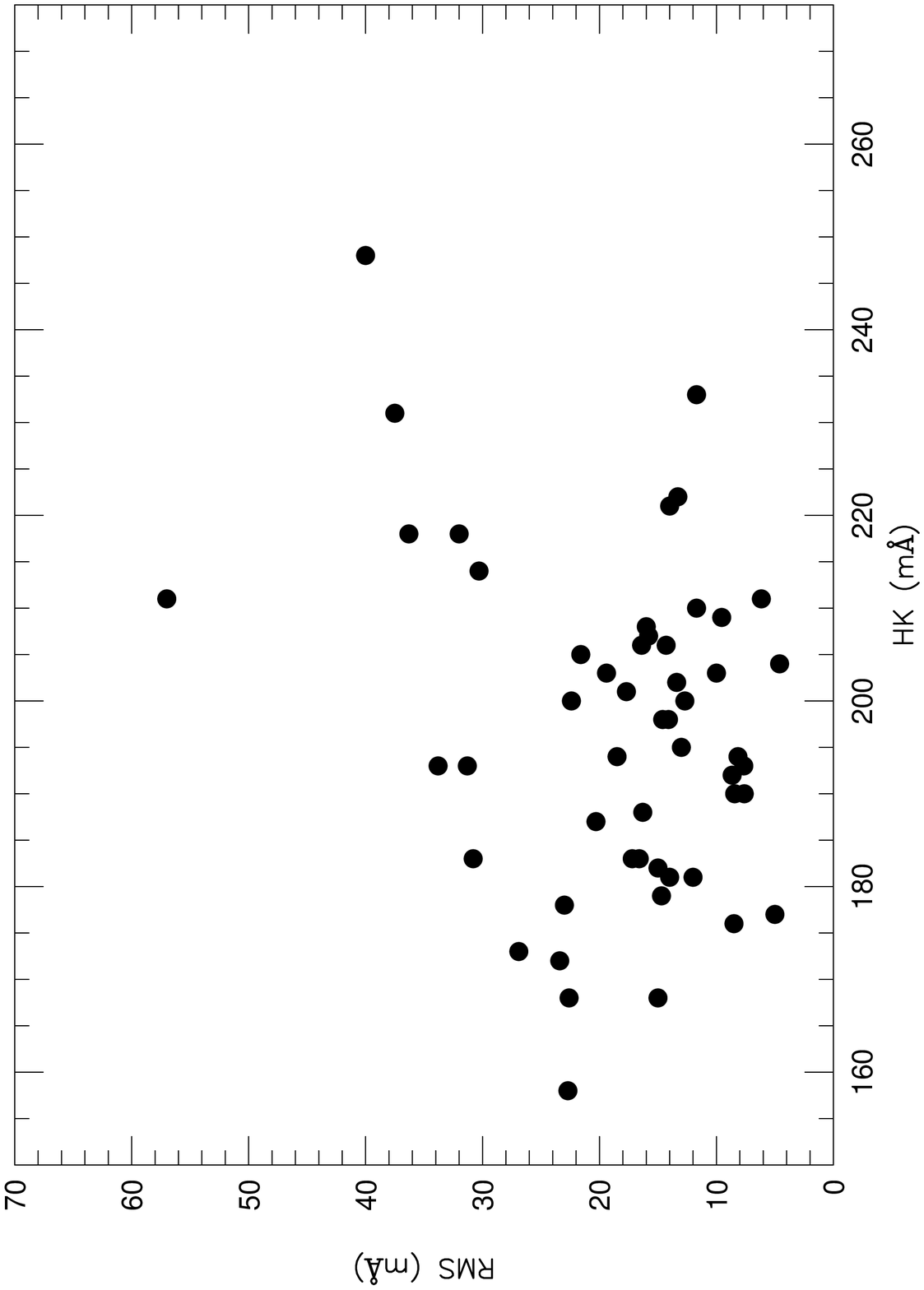}
\end{figure}

\begin{figure}
\figurenum{12}
\includegraphics*[scale=0.90,angle=180]{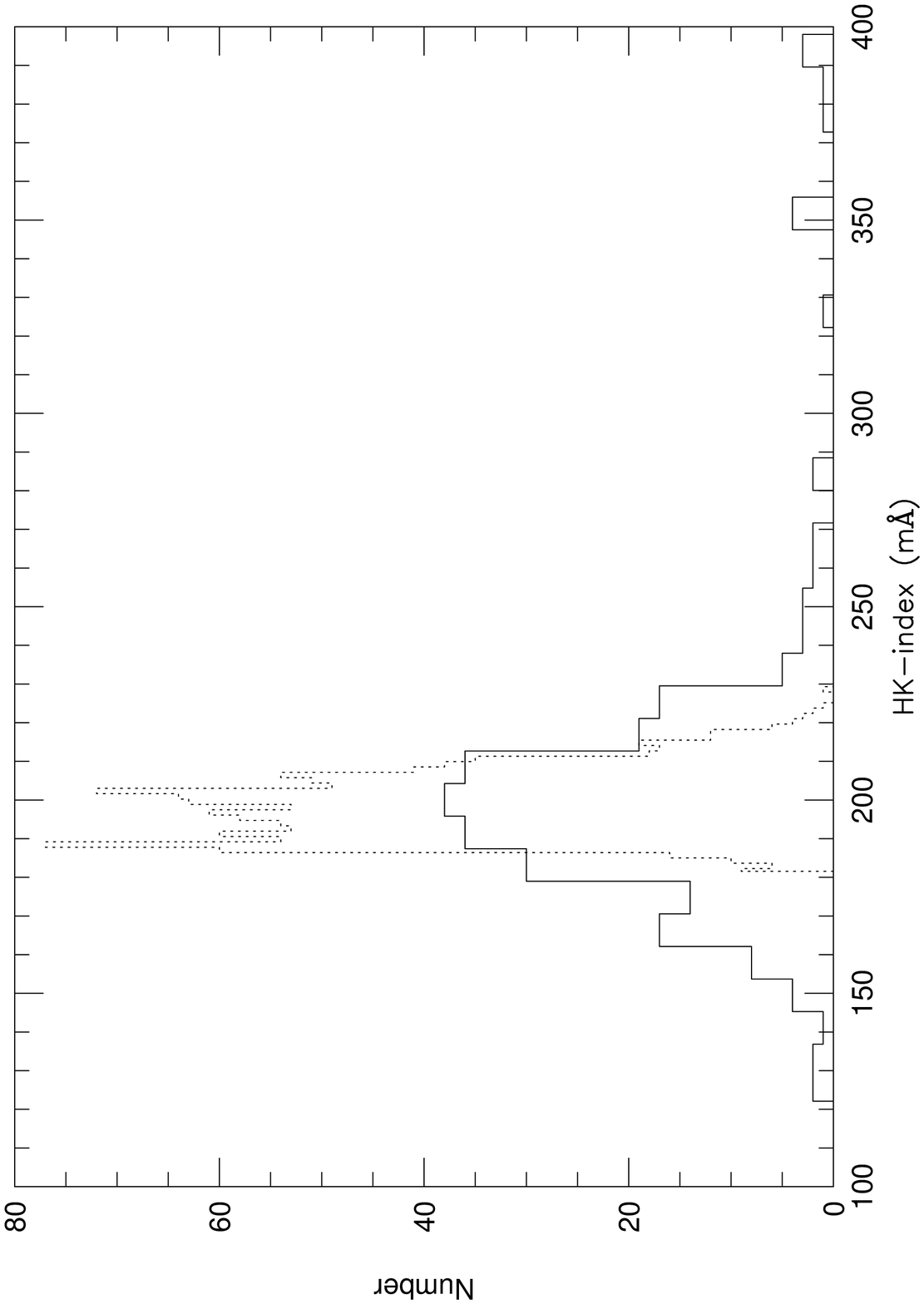}
\end{figure}

\begin{figure}
\figurenum{13}
\includegraphics*[scale=0.90,angle=180]{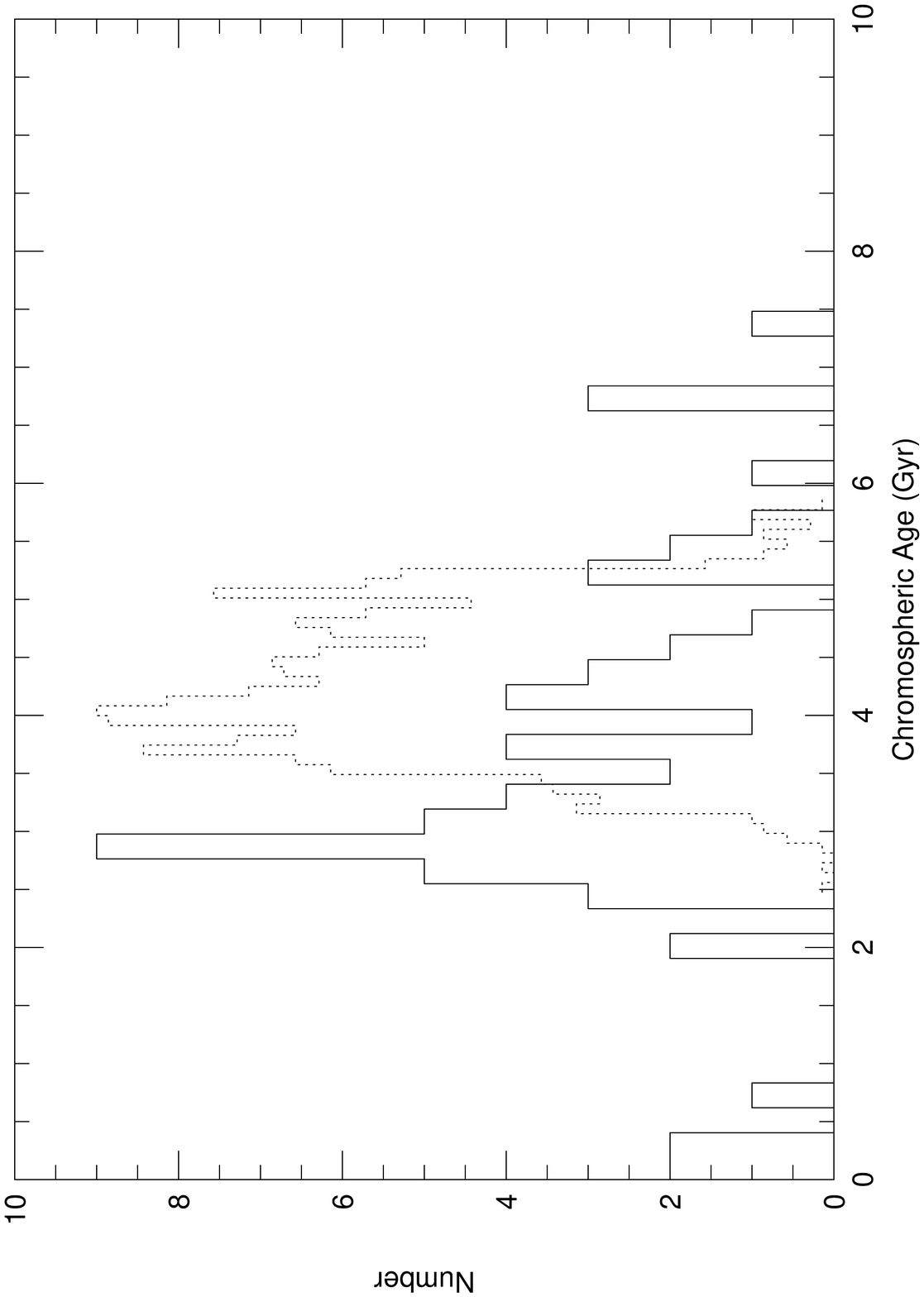}
\end{figure}

\begin{figure}
\figurenum{14}
\includegraphics*[scale=0.90,angle=180]{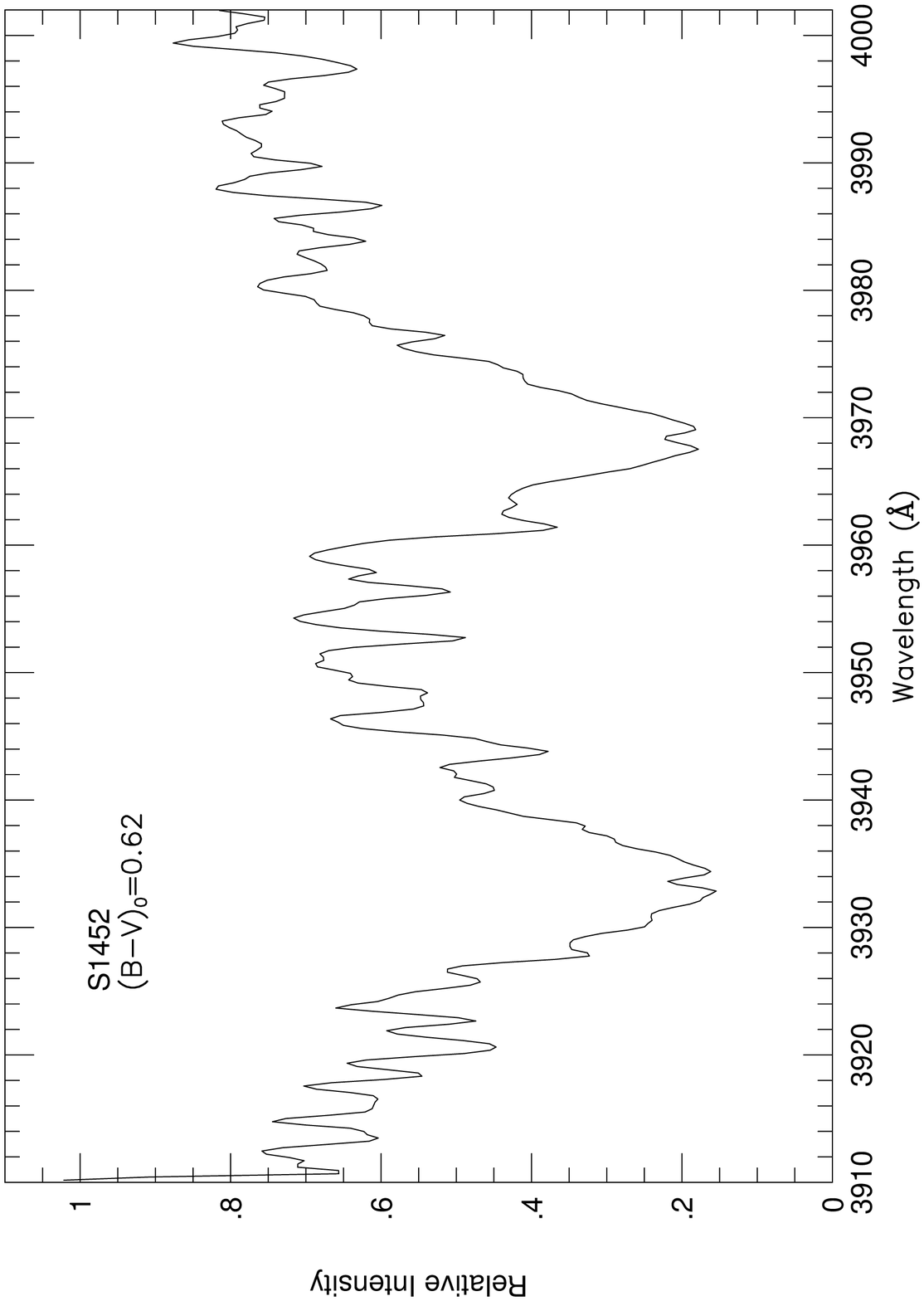}
\end{figure}

\end{document}